\documentclass{jfm}
\usepackage{xcolor}
\usepackage{natbib}
\usepackage{caption}
\usepackage{subcaption}
\usepackage{amsmath}
\usepackage[utf8]{inputenc}
\usepackage{MR_JFM}
\usepackage{graphicx}
\graphicspath{{fig_new/}}

\newcommand\calo[1]{{\cal{O}}(#1)}

\renewcommand{\Im}{{\rm Im}}

\def\so1{\sqrt{|\omega_1|}}

\def \ddpp#1#2{{\partial #1\over\partial #2}}
\def \ddpptwo#1#2{{\partial^2 #1\over\partial #2^2}}
\def \ddppthree#1#2{{\partial^3 #1\over\partial #2^3}}

\def\tauhat{\hat{\tau}}
\def\tautore{\tilde{\tau}}
\def\omegatore{\tilde{\omega}}
\def\abstauhat{|\hat{\tau}|}
\def\absomega1{|\hat{\omega}_1|}

\def\ymax{{y}_{\text{max}}}
\def\yattr{{y}_{\text{attr}}}

\def\euntiers{E^{1/3}}

\def\euntiers{E^{1/3}}
\def\emoinsuntiers{E^{-1/3}}

\def\newy{\bar{y}}
\def\ddpphy{\left.\ddpp{h}{y}\right\vert_{y}}

\title{Axisymmetric inertial modes in a spherical shell at low Ekman numbers}

\author[M. Rieutord, L. Valdettaro]{
M. RIEUTORD$^{1,2}$, L. VALDETTARO$^{3}$
}

\affiliation{
$^1$Universit\'e de Toulouse; UPS-OMP; IRAP; Toulouse, France\\
$^2$CNRS; IRAP; 14, avenue Edouard Belin, F-31400 Toulouse, France\\
[\affilskip]
$^3$MOX, Dipartimento di Matematica, Politecnico di Milano, Piazza L.
da Vinci, 32, 20133 Milano, Italy
}

\pubyear{2014}
\volume{000}
\pagerange{000--000}
\date{\today}
\begin{document}

\bibliographystyle{jfm}

\maketitle
%
%

\begin{abstract}

We investigate the asymptotic properties of axisymmetric inertial modes
propagating in a spherical shell when viscosity tends to zero. We identify
three kinds of eigenmodes whose eigenvalues follow very different
laws as the Ekman number $E$ becomes very small. First are modes
associated with attractors of characteristics that are made of thin
shear layers closely following the periodic orbit traced by the
characteristic attractor. Second are modes made of shear layers that
connect the critical latitude singularities of the two hemispheres of
the inner boundary of the spherical shell. Third are quasi-regular modes
associated with the frequency of neutral periodic orbits of
characteristics. We thoroughly analyse a subset of attractor modes for
which numerical solutions point to an asymptotic law governing the
eigenvalues. We show that three length scales proportional to
$E^{1/6}$, $E^{1/4}$ and $E^{1/3}$ control the shape of the shear layers
that are associated with these modes. These scales point out the key
role of the small parameter $E^{1/12}$ in these oscillatory flows. With
a simplified model of the viscous Poincar\'e equation, we can give an
approximate analytical formula that reproduces the velocity field in such
shear layers.  Finally, we also present an analysis of the quasi-regular
modes whose frequencies are close to $\sin(\pi/4)$ and explain why a
fluid inside a spherical shell cannot respond to any periodic forcing
at this frequency when viscosity vanishes.

\end{abstract}

\section{Introduction}

Oscillations of rotating fluids have long been a focus of fluid
mechanics. They are usually referred to as inertial oscillations but other
name may be used when more specific cases are considered (Kelvin waves,
Rossby waves, etc).  The first results in this field are due to the work
of \cite{kelvin1880} who gave the spectrum of the eigen oscillations of
a fluid rotating in an infinitely long cylinder.  This work was soon
followed by those of \cite{poinc1885} and \cite{bryan1889} who were
motivated by the stability of self-gravitating rotating ellipsoids for
their applications to planets and stars.

Presently, the motivations for studying oscillations of rotating fluids
are still vivid because of their implications in the understanding of
natural objects like stars, planets, oceans or the atmosphere of the
Earth, etc. Indeed, the observations of these oscillations in stars or
planets may readily give strong constraints on the global or
differential rotation of these bodies \cite[][]{RGV00,BR13}. Moreover,
as these oscillations are in the low-frequency range of the spectrum,
they are prone to tidal excitation and may play a crucial part in the
dynamical evolution of binary stars or close-in planets
\cite[][]{OL04,O09,RV10}. Similarly, the dynamics of a precessing
planets including internal fluid layers is also influenced by these
oscillations \cite[][]{HK95,Noir2001a}.

Closer to us, the dynamics of the oceans has also motivated many studies
of these modes but in cartesian geometry rather than in spherical
geometry \cite[e.g.][]{MM03}. Here, but this is also true in some stars,
inertial waves are part of the set of low frequency waves where we also
find internal gravity waves. These latter waves share many similarities
with inertial waves and often combine with them to form gravito-inertial
waves \cite[][]{FS82a}. As in stars or planets, oceanic internal
waves (but they are also found in the atmosphere), are sources of
dissipation, mixing, and momentum fluxes \cite[][]{Gerk08}.

Beyond the many applications that have been briefly mentioned,
studying waves propagating over rotating fluids is also motivated by the
mathematical problem that controls the dynamics of these flows. Indeed,
the linear equations that govern the small amplitude oscillations lead
to a mathematically ill-posed problem. If we consider the simplest case of
an inviscid, incompressible rotating fluid, small amplitude oscillations
of the pressure $p$ obey the Poincar\'e equation, namely

\beq \Delta p - \frac{4\Omega^2}{\omega^2}\ddz{p} = 0 \eeqn{poincare}
where $\Delta$ denotes the Laplacian operator,
$\vO=\Omega\ez$ is the angular velocity of the fluid and $\omega$
the angular frequency of the oscillation. 
As we pointed out, the oscillations
are low-frequency and one may easily show that $\omega\leq 2\Omega$
\cite[e.g.][]{Green69}. Hence, Poincaré operator is of hyperbolic type
leading to an ill-posed problem when associated with boundary
conditions. In such a case singularies are expected. The pathological
nature of these oscillations has soon been suspected \cite[][]{SR69} but
the clear evidence of the singularities had to await precise numerical
solutions to be exhibited \cite[][]{RV97}. Surprisingly, in some
containers like the infinitly long cylinder
\cite[][]{kelvin1880} or the ellipsoid \cite[][]{bryan1889}, analytical
solutions exist.

Recently, the completeness of the set of inertial modes as basis vector
functions for flows in some containers has been demonstrated. After the
pioneering work of \cite{cui_etal14} on the rotating annulus, \cite{IJW15}
have shown that the set of Poincar\'e modes is even complete in the
sphere, a result that is part of a more general one by \cite{backus15},
who also show that Poincaré modes indeed form a complete basis in the
ellipsoid. These results actually help understand the result of
\cite{ZELB01} showing the orthogonality of the inertial modes and
their associated viscous force.

The pathological nature of inertial modes is shared by internal gravity
modes \cite[][]{ML95} as they are also governed, in the inviscid limit
at the Boussinesq approximation, by the Poincaré equation.
Not unexpectedly, singularities also appear in the gravito-inertial modes
\cite[][]{DRV99}. In this latter case, the mathematical
nature of the underlying inviscid equations is a mixed-type operator
\cite[][]{Friedl82}, a property that is also found if the fluid is just
differentially rotating (without any stratification - e.g.
\citealt{BR13}). 

In many cases singularities appear because characteristics associated
with the hyperbolic problem get focused towards an attractor that
can be either a periodic orbit in a meridional plane or a wedge
made by boundaries or critical surfaces. When viscosity is included,
singularities are regularized. Those associated with periodic orbits
are transformed into detached shear layers. Such shear layers have been
observed experimentally by \cite{MBSL97} with pure gravity modes and by
\cite{MM03} for pure inertial modes.

\cite{HK95}, \cite{RV97} and \cite{RGV01} have shown that in the limit of
small viscosities, these shear layers seem to follow some asymptotic scaling
laws as far as their thickness is concerned. We are not affirmative
since no general demonstration exist. \cite{RVG02} have shown that in
two dimensions, namely in the meridional plane of a container but far
from the rotation axis so that curvature terms can be dismissed,
shear layers scale like $\nu^{1/4}$, where $\nu$ is the kinematic
viscosity of the fluid. But this result is specific to the restricted
2D-problem.

In the present work, we reconsider the set-up of a slightly viscous
rotating fluid inside a spherical shell as in \cite{RGV01} and investigate
the asymptotic properties of singular inertial modes. We only focus on
axisymmetric modes since non-axisymmetric inertial modes share the same
singularities, but probably in a milder way. Indeed, the trend of
non-axisymmetric modes to be closer to the outer boundary makes them
less sensitive to the presence of the core, which is the source of
singularities. Axisymmetric modes are in our opinion the best candidates
for deciphering the role of singularities in the modes dynamics.

In this study, we wish to understand the way
eigenfrequencies are determined and quantized around a given attractor
of characteristics and thus wish to generalize the work of \cite{RVG02}
to the associated three-dimensional system.  We shall see that the move
to three dimensions of space strongly affects the scaling laws and makes
the problem of much greater difficulty.

While re-investigating the properties of inertial modes in a spherical
shell at small viscosities, we can identify some robust scaling laws and
length scales in the shear layers, but the general solution or even the
quantization condition of a particular set of modes is still escaping
our understanding. Our results nevertheless delineate some interesting
properties of the modes that may help future work to finally circumvent
the difficulty of this problem and give the equation controlling the
structure of the shear layers and the associated quantization of the
eigenvalues.

The paper is organized as follows. In the next section we formulate the
mathematical problem and present the numerical method that is used.
Then, we present a set of numerical results that show clearly three
distinct sets of of eigenmodes. In section 4, we propose a first
analysis of the dynamics of shear layers associated with periodic
attractors of characteristics and show that we can recover the shape of
the eigenmodes but without any condition of quantization. A discussion
and some conclusions end the paper.

\section{Formulation of the problem}

\subsection{Equations of motion and boundary conditions}

We consider an incompressible viscous fluid inside a rotating spherical
shell of outer radius $R$ and inner radius $\eta R$ with $\eta<1$. Over
this solid body rotation at angular velocity $\Omega$, some small
amplitude perturbations propagate. If we use $(2\Omega)^{-1}$ as the
time scale and $R$ as the length scale, small-amplitude disturbances
obey the following non-dimensional linear equations:

\greq
\dt{\vu} + \ez \times \vu = -\na P + \EK\Delta\vu\\
\\
\na\cdot\:\vu=0
\egreqn{eqmo}
where 

\beq \EK = \frac{\nu}{2\Omega R^2} \eeqn{ekn}
is the Ekman number. We also introduced the unit vector along the rotation
axis $\ez$, and the pressure perturbation $P$. As we shall focus on
the eigenmodes of this system we impose perturbations to be proportional
to $\exp(\lambda t)$ where $\lambda$ is the complex eigenvalue. System
\eq{eqmo} needs to be completed by boundary conditions. We
impose impenetrable conditions ($\er\cdot\vu$) in the radial direction. 
In tangential directions we choose to impose stress-free conditions, namely,

\[ \er\times([\sigma]\er) = \vzero\]
where $[\sigma]$ is the non-dimensional viscous stress tensor and $\er$
the unit radial vector \cite[for an expression of this tensor see][for
instance]{rieutord15}. The choice of these boundary conditions is
not crucial \cite[e.g.][]{FH98}, but stress-free conditions are less
demanding on numerical resolution than the no-slip ones.

\subsection{Numerical method}

As in \cite{RGV01}, we discretize the partial differential equations
using a spectral decomposition. Namely, we expand the functions on the
spherical harmonics

\[\vu=\sum_{l=0}^{+\infty}\sum_{m=-l}^{+l}\ulm(r)\RL+\vlm(r)\SL+\wlm(r)\TL
,\]
with

\[\RL=\YL(\theta,\varphi)\vec{e}_{r},\qquad \SL=\na\YL,\qquad
\TL=\na\times\RL \]
where gradients are taken on the unit sphere. We then project the curl
of the momentum equation on the same basis and following \cite{R87},
we find

\greq
E\Deltal\wl - \lambda \wl= \\
 \hspace{1.5cm} -A_{\ell}r^{\ell-1}\dr{} \biggl( \frac{\ulmm}{r^{\ell-2}}\biggr)
-A_{\ell+1}r^{-\ell-2}\dr{}\biggl( r^{\ell+3}\ulp\biggr) \\
\\
E\Deltal\Deltal(r\ul)-\lambda \Deltal(r\ul)= \\
\hspace{1.5cm} B_{\ell}r^{\ell-1}\frac{\partial}{\partial r}
\biggl(\frac{\wlmm}{r^{\ell-1}}\biggr) + B_{\ell+1}r^{-\ell-2}
\dr{}\biggl( r^{\ell+2}w^{\ell+1}\biggr)
\egreqn{eqproj}
where axisymmetry has been assumed. We also introduced

\[ A_\ell = \frac{1}{\ell\sqrt{4\ell^2-1}}, \qquad
B_\ell = \ell^2(\ell^2-1)A_\ell, \qquad \Deltal = \frac{1}{r}\ddnr{}r -
\frac{\ell(\ell+1)}{r^2} \; .\]
where $\Deltal$ is the radial part of the scalar Laplacian
\cite[e.g.][]{R87}. Stress-free boundary conditions impose that

\[ \ul = \ddr{r\ul} = \dr{}\lp\frac{\wl}{r}\rp =0 \]
at $r=\eta$ or $r=1$ for the radial functions.

System \eq{eqproj} is then discretized on
the collocation points of the Gauss-Lobatto grid. Including boundary
conditions, the system can be written as a generalized eigenvalue
problem like

\[ [A]\vX = \lambda [B]\vX\]
where $[A]$ and $[B]$ are matrices whose dimension depends on the
numerical resolution.  We are mostly interested in the least-stable
eigenmodes, which are associated with the generalized eigenvalues
$\lambda$ with the greatest real part. We solve this problem using the
incomplete Arnoldi-Chebyshev method (\cite{CHAT12,VRBF07}). Let $\mu$
be the solutions of the modified problem

\beq
([A]-\sigma [B])^{-1} [B] X = \mu X.
\label{eq:sigma}
\eeq

Then $\lambda=\sigma+1/\mu$. Thanks to this transformation, the
eigenvalues near the shift (the guess) $\sigma$ are the extreme
eigenvalues of this modified problem and are thus delivered by the
Arnoldi procedure.  Nowadays machines allow us to find eigenvalues with
matrices of order up to $5\times10^6$ corresponding to the use of 3000
spherical harmonics and 1500 radial grid points using double precision
arithmetic. All numerical solutions presented below own in general a
relative truncation error for the eigenfunctions less than $10^{-3}$,
which is achieved by the resolution indicated by the $L_{\rm max}$
and $N_r$ values.

\begin{figure} \centering
\centerline{\includegraphics[width=0.49\linewidth]{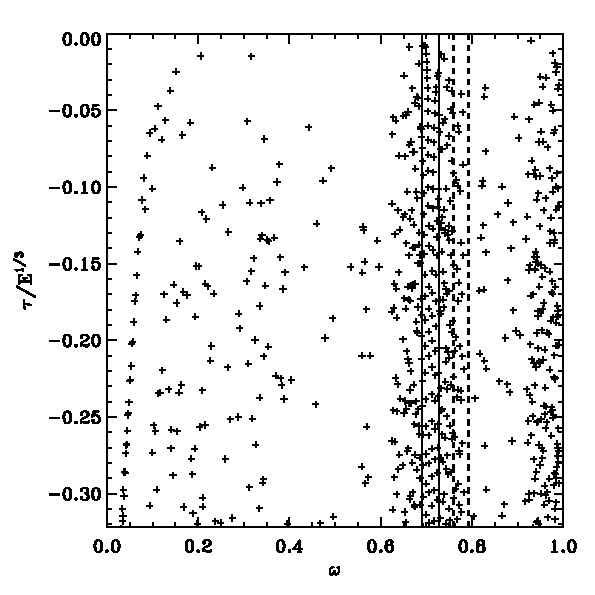}
\includegraphics[width=0.49\linewidth]{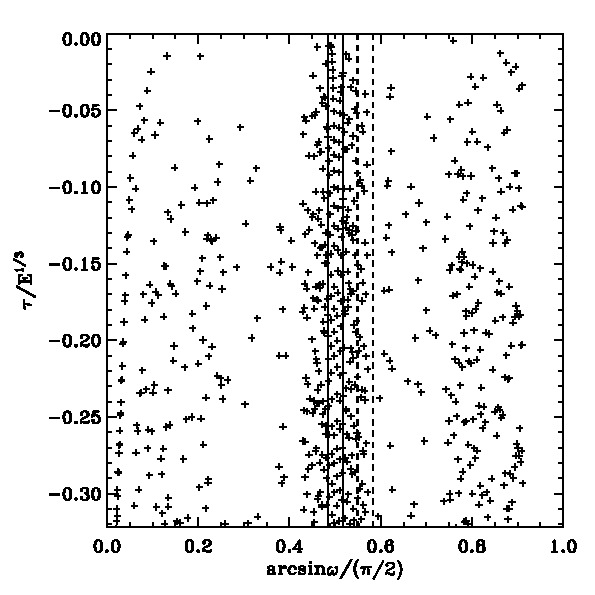}}
\caption[]{Left: Distribution of eigenvalues of inertial modes in the complex
plane. Right: same as left but the frequency has been converted into an
angle (the critical latitude) scaled by $\pi/2$. The dashed vertical
lines delineate the part of the complex plane that is magnified in
figure~\ref{spec782} while the solid vertical lines outline the part of the
complex plane shown in figure~\ref{sp_fontain1}a. The Ekman number is
set to 10$^{-8}$, $\eta=0.35$ and the numerical resolution is $L_{\rm
max}=800$ and $N_r=300$. In both figures the real part has been rescaled
by $E^{1/3}$.}
\label{fullspect}
\end{figure}

\section{Numerical results}

The numerical investigation  of the foregoing eigenvalue problem that
we shall now present, has revealed several types of eigenmodes.

We shall restrict in the following to modes that are symmetric with respect to the equator. 
The classification is based on the path of characteristics associated
with the Poincaré equation. We recall that system \eq{eqmo} can be reduced
to a single equation for the pressure perturbation, namely \eq{poincare}
when viscosity is set to zero. In the dimensionless expression of the
equation, the frequency of the oscillation $\omega$
is necessarily less than unity.  Associated characteristic
surfaces are cones (or parts of cones) characterized by their opening
angle $\theta^c=\arcsin\omega$ and an apex on the rotation
axis. $\theta^c$ is also the critical latitude. This is the latitude
where the characteristics surfaces are tangent to the spheres.  Even for
non-axisymmetric modes, characteristic surfaces are axisymmetric cones
\cite[][]{RGV01}. This is why we shall always visualize the characteristic
cones by their trace in a meridian plane where they appear as straight
lines. As \cite{RGV01} have shown, the path of the characteristic
lines in a meridian plane generally converges towards a closed periodic
orbit that is called an attractor.  Exceptions are a finite number of
frequencies that read $\sin(p\pi/q)$ where $p$ and $q$ are integers. For
these frequencies any trajectory is periodic and there is no attractor. The
number of such frequencies depends on the aspect ratio of the shell. For
$\eta=0.35$, periodic orbits with $p=1$ 
exist only for $q=3,4,6,8$ \cite[][]{RGV01}.

To set the stage, we show in figure~\ref{fullspect} a general view of
the distribution of eigenvalues associated with viscous inertial modes
in the complex plane. These eigenvalues have been computed through a
systematic scan of the least-damped part of the complex plane with the
Arnoldi-Chebyshev algorithm. It extends figure~17 of \cite{RGV01}. When
the imaginary part of the eigenvalues (the frequency) is converted
into an angle (actually the critical latitude), the approximate symmetry with
respect to latitude $\pi/4$ is emphasized. This symmetry is verified by
characteristics trajectories, but not by the eigenfunctions since the
rotation axis is of course not the same as the equator. The distribution
of eigenvalues reflects this symmetry near $\pi/4$, but this symmetry
weakens when the critical latitude of the modes moves away from $\pi/4$.

This general view of the complex plane clearly shows that the
distribution is not uniform and no simple quantization, or quantum
numbers, controls it. However, some regularities appear: A crowded
region near $\pi/4$ (quasi-regular modes), deserted regions around $\pi/6$
and $\pi/3$, and some deep frequency bands where modes are strongly
damped.

As shown by \cite{RGV01}, this distribution of eigenvalues is profoundly
marked by the orbits of characteristics and the attractors they may
form. We have found three categories of modes, which we termed as
attractor modes, critical latitude modes and quasi-regular modes
respectively.

\subsection{Attractor modes}

Attractor modes are modes associated with
a specific attractor represented by a periodic orbit of characteristics.
These modes were first studied in \cite{RV97}, and their analytic
expression has been given by \cite{RVG02} in the two-dimensional
case\footnote{We recall that the two-dimensional case refers to the same case
as the one described by equation \eq{eqmo} but where curvature terms
(like $\frac{1}{r}\dr{}$) of the spherical geometry are dismissed.
This case is also referred to as the case of the slender torus, which is
a torus with a large aspect ratio. \cite{RVG02} have shown that it can
describe inertial modes that are trapped in the equatorial region of a
thin spherical shell.}.

\begin{figure} \centering
   \begin{subfigure}{.48\textwidth} \vfill \centering
      \includegraphics[width=\linewidth,clip=true]{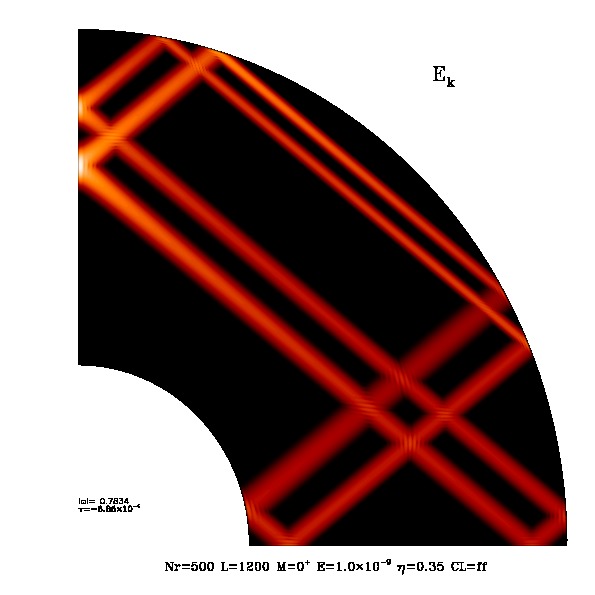}
      \caption[]{}
   \end{subfigure} \hfill
   \begin{subfigure}{.48\textwidth} \vfill \centering
      \includegraphics[width=\linewidth,clip=true]{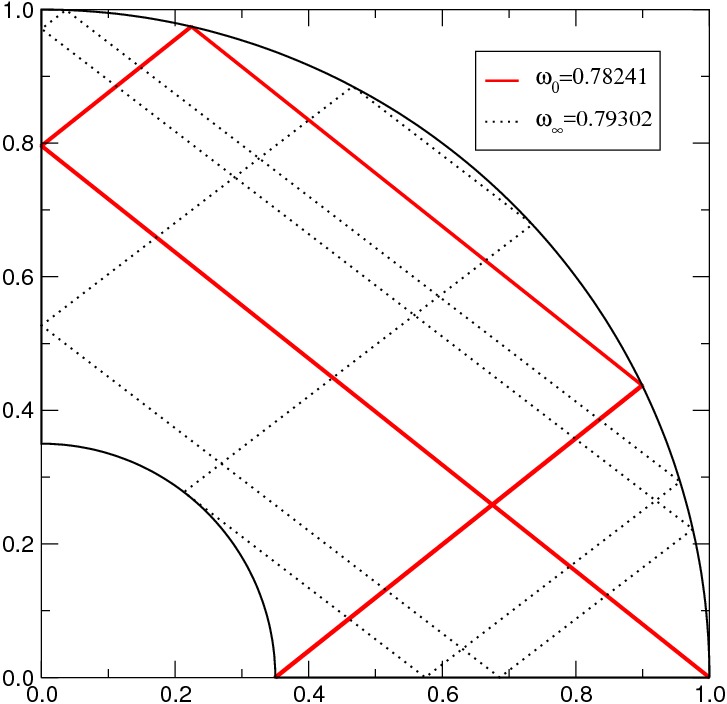}
      \caption[]{}
   \end{subfigure}
   \caption[]{
(a) The least-damped eigenmode associated with the attractor at
$\omega=0.7834$ and $E=10^{-9}$. The numerical resolution is $L_{\rm
max}=1200$, $N_r=500$.  (b) Red line: The asymptotic attractor $\omega_0\simeq
0.782413$ associated with the mode in (a).  Black-dotted line: The
same attractor at the upper frequency limit $\omega=\omega_\infty\simeq
0.793$.} \label{m782}
\end{figure}

\begin{figure} \centering
      \includegraphics[width=.7\linewidth,clip=true]{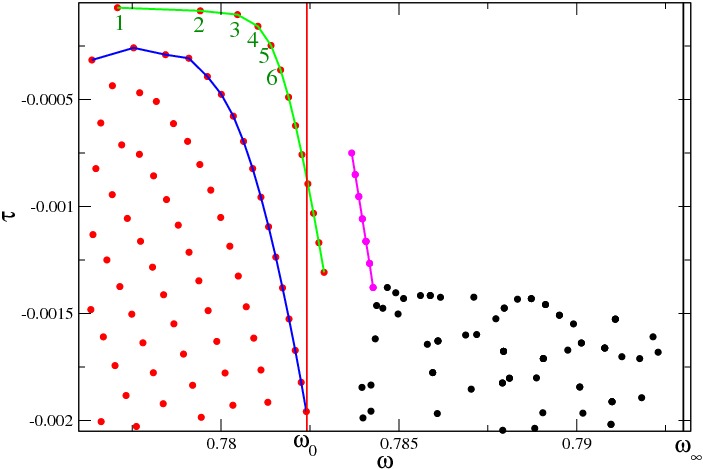}
\caption[]{Distribution of eigenvalues in the complex plane around
the frequency 0.785 for $\eta=0.35$ and E$=3\times10^{-9}$. The
corresponding attractor is shown in figure~\ref{m782}. $\omega_0$ and
$\omega_\infty$ are the frequency bounds of the attractor. Red dots show
eigenvalues associated with critical latitude modes while black dots
show eigenvalues associated with attractor modes. The purple eigenvalues
have been computed with extended precision.}
\label{spec782}
\end{figure}

In figure~\ref{m782}a we show one such attractor mode. Eigenmodes
featured by this attractor have eigenfrequencies in the interval
$[\omega_0, \omega_\infty]$ where $\omega_0\simeq 0.782413$ and
$\omega_\infty\simeq 0.793$ for $\eta=0.35$. In figure~\ref{m782}b, we
show the limiting shapes of this attractor when
when $\omega=\omega_0$ or $\omega=\omega_\infty$.
We recall that the strength of an attractor may be characterized
by a (negative) Lyapunov exponent that measures the rate at which
characteristics converge towards the attractor. $\omega_0$ and
$\omega_\infty$ refer to the values where the Lyapunov exponent is
respectively zero or $-\infty$. In the former case, characteristics
are still converging towards the attractor but algebraically, while
in the latter case they touch the critical latitude making the mapping
(featured by the characteristics) infinitely contracting \cite[see][for
a more detailed discussion]{RGV01}. Note that the asymptotic frequency
$\omega_0$ of Fig.~\ref{m782}-attractor is easily expressed as a function
of the aspect ratio $\eta$ and reads

\beq \omega_0=\sqrt{\frac{3+\sqrt{5-4 \eta}}{8}}\eeqn{omega0}
In figure~\ref{spec782}, we now show the part of the complex plane where
the eigenvalues of this attractor show up. The vertical lines delineate
the limiting frequencies $\omega_0$ and
$\omega_\infty$. We immediately note that eigenvalues are distributed in
several subsets. In the interval $[\omega_0, \omega_\infty]$, the purple
dots show the eigenvalues associated with the attractor modes, while the
black dots are affected by
numerical noise (see below).  At frequencies
lower than $\omega_0$, we note a neat organisation of the eigenvalues
(red dots), which is associated with the set of ``critical latitude
modes". We discuss these latter modes in the next section.

\begin{table}
\begin{center}
\begin{tabular}{ccccccccc}
$\omega_0$ & $\alpha_0$  & $\tauhat_1$ & $2\tauhat_1$ & $\phi_1$ & $\tauhat_2$  &
$\sqrt{2}\tauhat_2$ & $\phi_2$ & $\eta$\\
0.555369 & 0.831694 & 0.5085     & 1.017   &$+\pi/3$ & 2.275 & 3.217
& $+\pi/4$ & 0.35 \\
0.831694 & 0.555369 & 0.812      & 1.62     &$-\pi/3$ & 1.65 & 2.33
& $-\pi/4$ & 0.35 \\
0.622759 & 0.782413 & 0.565      & 1.13     &$+\pi/3$ & 2.1  & 2.97
& $+\pi/4$ & 0.35 \\
0.782413 & 0.622759 & 0.485      & 0.97     &$-\pi/3$ & 1.82 & 2.57
& $-\pi/4$ & 0.35 \\
0.466418 & 0.884564 & 0.485      & 0.97     &$+\pi/3$ & 3.95 & 5.586
& $+\pi/4$ & 0.50 \\
0.884564 & 0.466418 & 0.645      & 1.29     &$-\pi/3$ & 3.05 & 4.31
& $-\pi/4$ & 0.50 \\
\hline
\end{tabular}
\caption[]{Asymptotic parameters of six modes following \eq{egv_exp}.}
\label{am}
\end{center}
\end{table}

Remarkably, the eigenvalues of the attractor modes (the purple
dots in Fig.~\ref{spec782}) seem to be
governed by the following law:

\beq \lambda_n = i\omega_0- 2\tauhat_1 e^{i\phi_1}\EK^{1/3} -
\lp n+\demi\rp e^{i\phi_2} \sqrt{2}\tauhat_2\EK^{1/2}+\cdots
\eeqn{egv_exp}

\begin{figure} \centering
\centerline{\includegraphics[width=0.95\linewidth,clip=true]{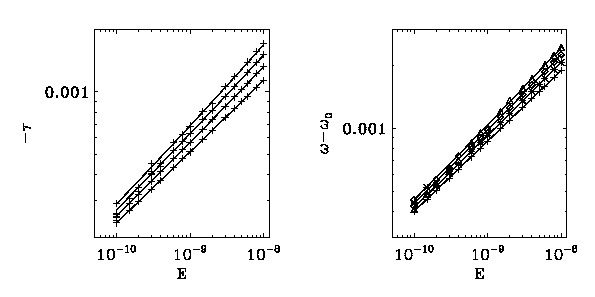}
}
\caption[]{Left: Damping rate for the 4 least-damped eigenvalues of
the 0.782-attractor modes as a function of the Ekman number. Pluses
indicate the numerical values and the solid lines show the law
\eq{egv_exp}. Right: same as left but for the deviation of the frequency
from the frequency of the asymptotic attractor.
}
\label{scal782}
\end{figure}

In this expression, $\omega_0$ is the asymptotic frequency of the
attractor (as given by Eq.~\ref{omega0}), while $\tauhat_1$ and $\tauhat_2$ are
positive real numbers of order unity that depend on the shape of the
attractor.  $n$ is the quantum number that characterises the mode. 
We use the term ``quantization" in this context to signify that
eigenvalues are arranged along specific lines in the complex plane and are
distributed with some regularity along such lines. We note that the
$n+1/2$ factor is reminiscent of the energy levels of a quantum particle
in a parabolic well, and of the eigenvalues of the two-dimensional
problem of \cite{RVG02}.  In figure~\ref{scal782}, we illustrate the
good matching of the real and imaginary parts of the eigenvalues with
the law~\eq{egv_exp}.

We have found such sets of eigenvalues in association with various
attractors. Table~\ref{am} gives the parameters for six sets of such
modes. In this table, we gathered the families of modes by pairs of
families where we associated the attractor at $\omega_0$ with the
symmetric one at $\alpha_0=\sqrt{1-\omega^2_0}$. The symmetry is with
respect to latitude $\pi/4$. We note that the phase $\phi_1$ or $\phi_2$
in ~\eq{egv_exp} are opposite for pairs of attractors.  This betrays
the fact that $\omega_\infty<\omega<\omega_0$ when the frequency is
less than $1/\sqrt{2}$ while $\omega_\infty>\omega>\omega_0$ when
$\omega>1/\sqrt{2}$.  Now, we observe that the symmetry of attractors,
with respect to latitude $\pi/4$, is not verified by the modes since the
$\tauhat_1$ and $\tauhat_2$ coefficients are not the same for a family and
its symmetric. The $\tauhat_1$ and $\tauhat_2$ coefficients are therefore
sensitive to the reflection on the rotation axis.  We remark that there
are attractors with  $\omega_0<\omega<\omega_\infty$
for frequencies less than $1/\sqrt{2}$ (like the ones with
$\omega_0=0.35866$ and $\omega_\infty=0.36134$) and correspondingly
$\omega_0>\omega>\omega_\infty$ when $\omega>1/\sqrt{2}$. However, we did
not find any set of eigenvalues associated with those attractors. To be
complete, we note that there exist attractor modes whose eigenvalues
cannot be represented by \eq{egv_exp}. Since no clear law seems to
govern their properties, we shall not discuss them any further in the
present work.

\begin{figure} \centering
\centerline{\includegraphics[width=0.49\linewidth]{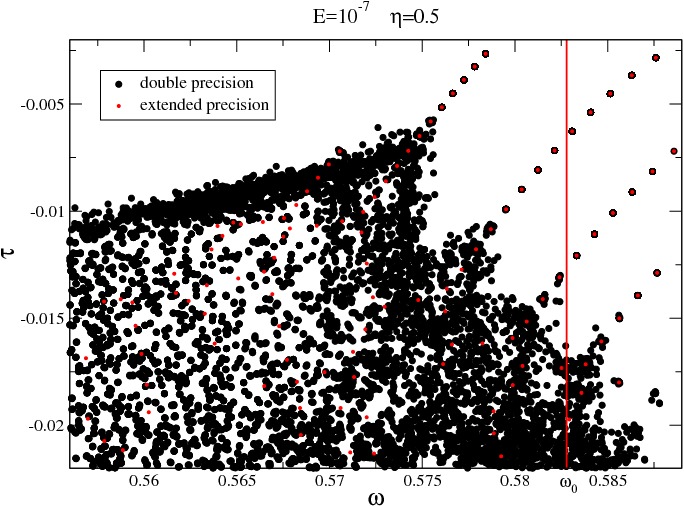}\hfill
\includegraphics[width=0.49\linewidth]{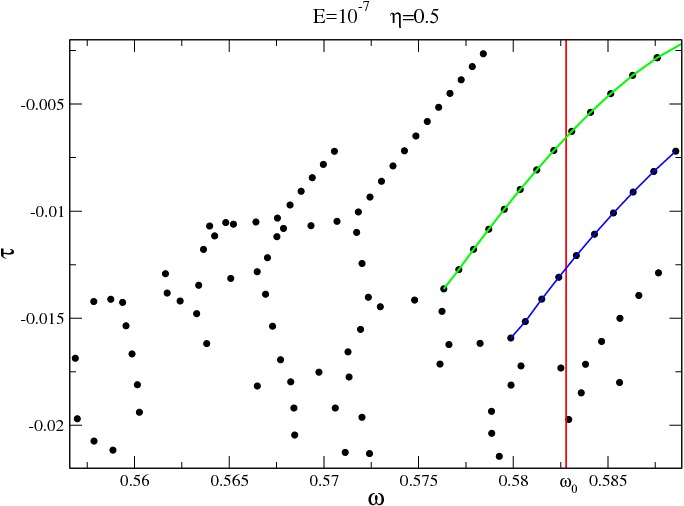}}
\caption[]{Eigenvalue spectrum of attractor-modes in the complex plane.
$\tau$ is the real part and $\omega$ is the imaginary part. Left: a
double precision calculation, right: with extended precision. In both
figures the resolution is $L_{\rm max}=440$ and $N_r=160$. The attractor
exists at frequencies below $\omega_0=0.583$ which is visualized by a
red line. The aspect ratio is $\eta=0.50$.  This attractor is the one
corresponding to the 0.555 when $\eta=0.35$ \cite[e.g.][]{RGV00}. The
green and blue lines emphasize two families of critical latitude modes
(see sect.~\ref{CLmodes}).
} \label{spec583}
\end{figure}

Round-off errors are actually a major plague of eigenvalue/eigenmode
computation of attractor modes. 
 We see in figure~\ref{spec782} that the black dots
associated with attractor modes are randomly distributed unlike
the least-damped modes (purple dots) which obey the dispersion relation
\eq{egv_exp}.  In fact, eigenvalues of the more damped modes
are strongly perturbed by round-off errors.

\begin{figure} \centering
   \begin{subfigure}{.48\textwidth} \vfill \centering
      \includegraphics[width=.9\linewidth,clip=true]{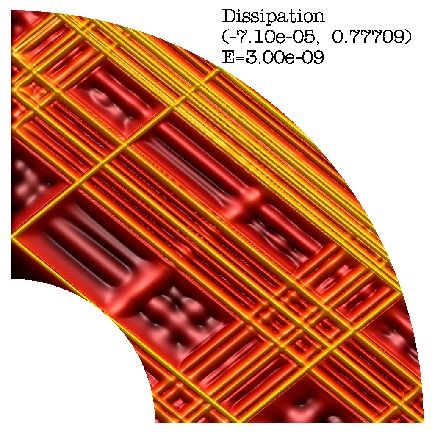}
      \caption[]{}
   \end{subfigure} \hfill
   \begin{subfigure}{.48\textwidth} \vfill \centering
      \includegraphics[width=.9\linewidth,clip=true]{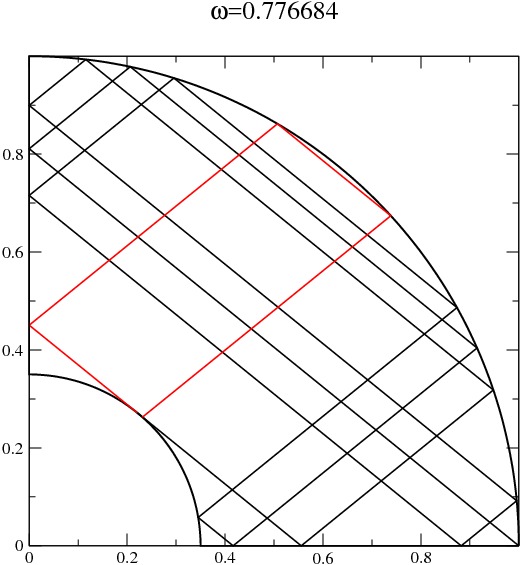}
      \caption[]{}
   \end{subfigure}
   \begin{subfigure}{.48\textwidth} \vfill \centering
      \includegraphics[width=.9\linewidth,clip=true]{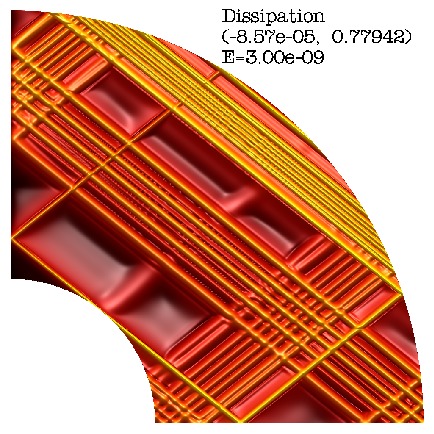}
      \caption[]{}
   \end{subfigure} \hfill
   \begin{subfigure}{.48\textwidth} \vfill \centering
      \includegraphics[width=.9\linewidth,clip=true]{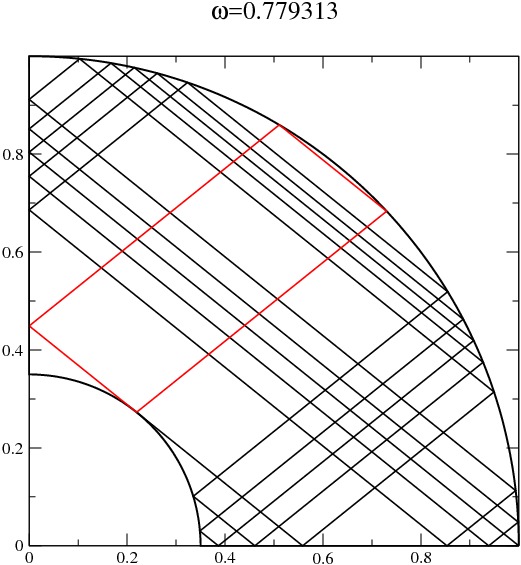}
      \caption[]{}
   \end{subfigure}
   \caption[]{
(a) and (c) : The viscous dissipation for the first and second eigenmodes
of the ``green family" shown in figure \ref{spec782}. The Ekman number is
$3\times10^{-9}$, for which we used $L_{\rm max}=1200$, $N_r=500$. Most
intense values are in yellow.  We also added an artificial depth (yellow
curves are on top) to better distinguish the shapes.  (b) and (d) show
the corresponding web of characteristics. The red rectangle shows the
periodic orbit drawn by the characteristic emitted towards the North
from the critical latitude.
}
   \label{latcrit}
\end{figure}
\begin{figure} \centering
   \begin{subfigure}{.48\textwidth} \vfill \centering
      \includegraphics[width=\linewidth,clip=true]{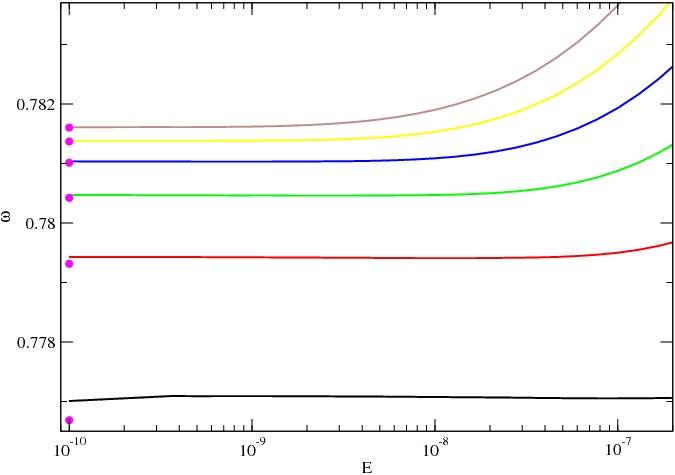}
      \caption[]{}
   \end{subfigure} \hfill
   \begin{subfigure}{.48\textwidth} \vfill \centering
      \includegraphics[width=\linewidth,clip=true]{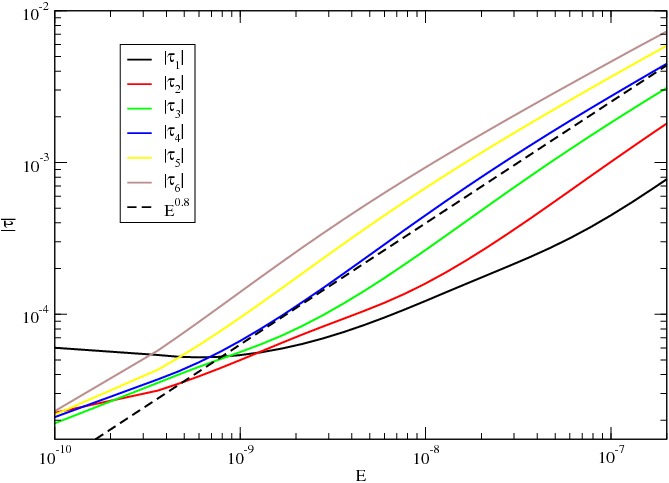}
      \caption[]{}
   \end{subfigure}
   \caption[]{
(a) Dependence of the frequency of the ``green family" modes of
Fig.~\ref{spec782} with the Ekman number. The pink dots  mark the
frequencies of the webs of rays that connect the North and South
critical latitudes of the internal sphere. (b) Same as in (a) but for
the damping rates. The numbers refer to the numbering of the mode in
figure~\ref{spec782}.
}
   \label{omegataulatcrit}
\end{figure}

In figure~\ref{spec583}, we further illustrate this phenomenon. The
left (resp. right) figure shows the double (resp. extended)
precision calculation of the distribution of eigenvalues associated
with attractor modes (for this attractor $\omega_\infty<\omega_0$).
The plotted eigenvalues are actually superpositions of several independent
calculations. In each calculation we have only changed the shift parameter
of equation \eq{eq:sigma}.  As was shown in \cite{VRBF07}, a noisy
distribution of eigenvalues is related to the sensitivity of matrices $[A]$
and $[B]$ to small perturbations, so ultimately to round-off errors. In
the left figure, we see that families governed by a dispersion relation
disappear in a bath of randomly distributed eigenvalues at large damping
rates. When the same distribution is computed with extended precision
(quadruple precision, right figure), the noisy distribution leaves the
place to an ordered distribution of eigenvalues. On this same figure
we also note that some branches cross the line marking the asymptotic
frequency $\omega_0$ of the attractor.  Examination of the modes along
this branch reveals that they are still featured by the asymptotic
attractor, despite the fact that the propagation of characteristics
does not show the attractor. The eigenfunctions whose eigenfrequency is
neatly above $\omega_0$ actually show that the modes of such a branch
are featured by the shear layer emitted towards North by the critical
latitude singularity. The associated characteristics trajectory shows
that the shear layer has to wind around the former attractor before
leaving it. However, because of viscosity, the winding stops at some
stage still leaving the trace of the attractor. Hence the branches can
continuously cross the $\omega_0$ line. We surmise that for asymptotically
small values of the Ekman number such crossing is not possible because
trajectories of characteristics may bifurcate towards another attractor.
In the next subsection we shall investigate such branches of modes.

\subsection{Modes associated with the critical latitude of the inner
sphere}\label{CLmodes}

Beside the modes that are associated with a periodic attractor, the
spectra (figure~\ref{spec782} and \ref{spec583}) display other obvious
families of modes. In figures~\ref{latcrit}(a) and \ref{latcrit}(c),
we show the two modes numbered 1 and 2 of the green-family of
figure~\ref{spec782}, with their associated path of characteristics
(Fig. \ref{latcrit}b and \ref{latcrit}d).  With these meridional
cuts, we clearly see that the characteristic emitted by the northern
critical latitude of the inner shell in the southern  direction finally
reaches the equator of the outer shell. It means, by symmetry, that
it joins the southern critical latitude on the inner shell. The shear
layer issued from the northern critical latitude towards the North
loops back to the same critical latitude as shown by the red path in
Fig.~\ref{latcrit}.

From the plots of Fig.~\ref{latcrit}, we note that the quantization of
these modes seems to come from the length of the path connecting the North and
South critical latitudes on the inner sphere. The path is not unique and
a slight change in the frequency increases the number of rays in the bulk
by two units. In Fig.~\ref{omegataulatcrit}a, we show that the discrete
frequencies of the green-family can almost be retrieved by the simple
geometrical rule of finding a path of characteristics that connects the
North and South critical latitudes. The same is true for the
blue family, which is characterized by more dissipative shear layers
(the transverse wavenumber is higher). The evolution of the damping
rate of these modes with the Ekman number is not standard as may be
seen in Fig.~\ref{omegataulatcrit}b. Indeed, for the first mode
(tagged 1 in Fig.~\ref{spec782}), below E=$10^{-9}$, the damping rate
increases while the Ekman number decreases. We explain this behaviour
as a consequence of the activation of the northern branch of the shear
layer that loops back to the critical latitude 
(red segments in Fig.~\ref{latcrit}), which is a place of
high dissipation. If the Ekman number is low enough, this loop has a
larger amplitude and has a larger contribution to the damping rate of
the mode. Most likely, ``critical latitude modes" are not asymptotic
and may only exist in a finite range of Ekman numbers. 

\begin{figure} \centering
   \begin{subfigure}{.48\textwidth} \vfill \centering
      \includegraphics[width=\linewidth,clip=true]{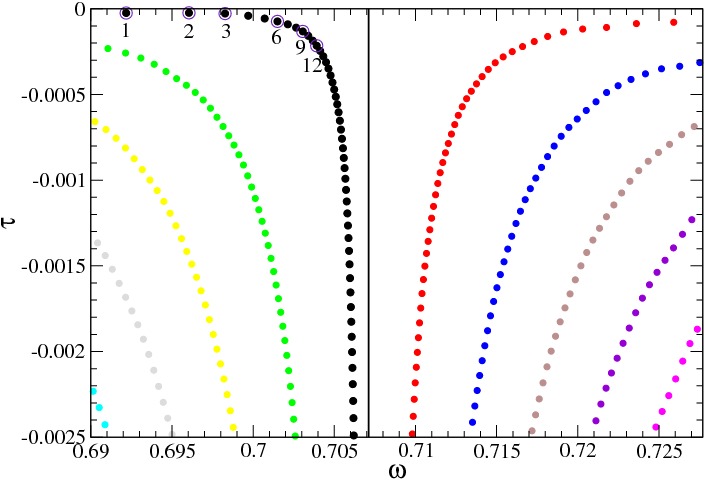}
      \caption[]{}
   \end{subfigure} \hfill
   \begin{subfigure}{.48\textwidth} \vfill \centering
      \includegraphics[width=\linewidth,clip=true]{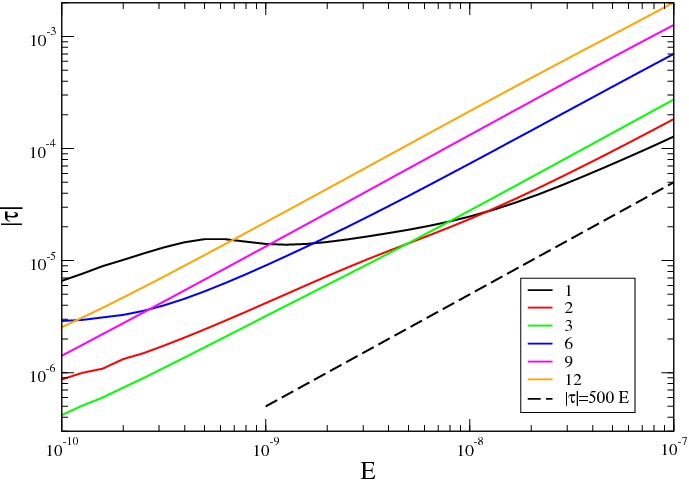}
      \caption[]{}
   \end{subfigure}
   \caption[]{
(a) Spectrum near the frequency $\sin(\pi/4)$. (b) Damping rate
as a function of $E$ for the modes numbered $1, 2, 3, 6, 9, 12$ in (a).
The dashed line is $|\tau|=500 E$.
}
   \label{sp_fontain1}
\end{figure}

\begin{figure} \centering
   \begin{subfigure}{.48\textwidth} \vfill \centering
      \includegraphics[width=\linewidth,clip=true]{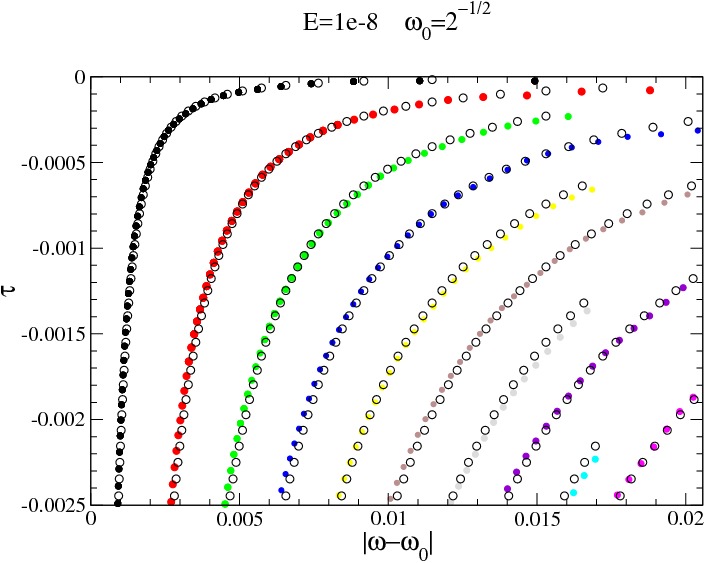}
      \caption[]{}
   \end{subfigure} \hfill
   \begin{subfigure}{.48\textwidth} \vfill \centering
   \includegraphics[width=\linewidth,clip=true]{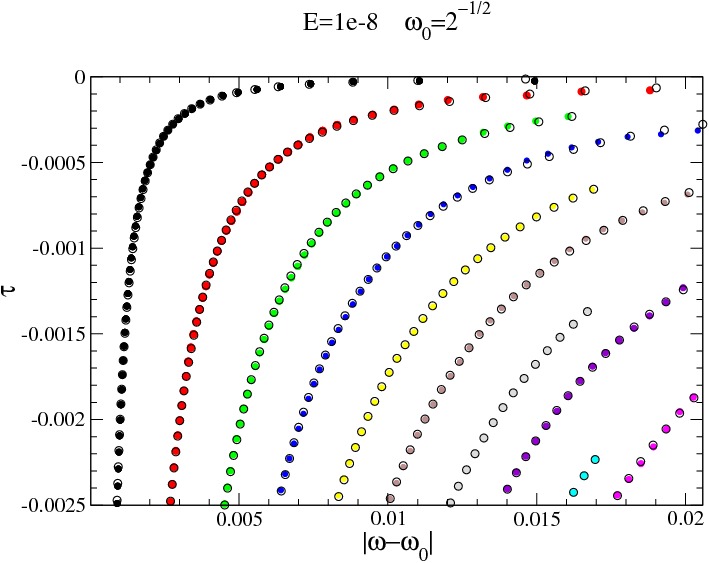}
      \caption[]{}
   \end{subfigure}
   \caption[]{
Distribution of eigenvalues (coloured dots) in the plane
$\tau,|\omega-\sin(\pi/4)|$.  (a) Open circles are obtained
using formula (\ref{eq:modes_reguliers_formule_simple} with
$a=0.141$, $b=0.152$).  (b) Open circles are obtained using formula
(\ref{eq:modes_reguliers_formule_compliquee}).
}
   \label{repliee}
\end{figure}

\subsection{The quasi-regular modes}

In the distribution of eigenvalues shown in Fig.~\ref{fullspect}, we
noticed a set of eigenvalues with very low damping rates gathered around
the frequency $\sin(\pi/4)$. A close up view of this region of the
complex plane, displayed in figure~\ref{sp_fontain1}a, shows that this set
of eigenvalues has peculiar properties that deserve some attention.
First, we note that the eigenvalues seem to obey simple quantization
rules as their distribution clearly follows specific lines in the
complex plane. In addition, their damping rate is almost proportional to
the Ekman number in some range of this parameter (e.g.
figure~\ref{sp_fontain1}b). These features give evidence of a
quasi-regular nature of this kind of modes. We recall that regular
eigenmodes have a structure that is weakly dependent on viscosity and
which converges to a smooth eigenfunction in the inviscid limit. We thus call
these modes quasi-regular since they are similar to truly regular modes in
some Ekman number range, but they lose this character below some Ekman
number specific to the mode (see below).

The specific distribution of eigenvalues in this region of the complex
plane can be explained with some simple arguments based on the
properties of the web of characteristics.

\begin{figure} \centering
   \begin{subfigure}{.49\textwidth} \vfill \centering
      \includegraphics[width=\linewidth,clip=true]{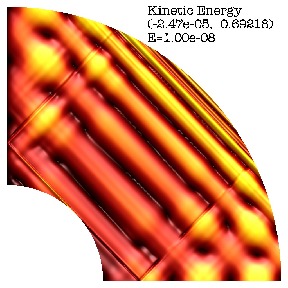}
      \caption[]{}
   \end{subfigure} \hfill
   \begin{subfigure}{.49\textwidth} \vfill \centering
      \includegraphics[width=\linewidth,clip=true]{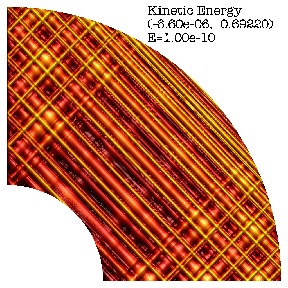}
      \caption[]{}
   \end{subfigure} \hfill

   \caption[]{(a) Meridional distribution of the kinetic energy of the
   eigenmode associated with eigenvalue tagged ``1" in
Fig.~\ref{sp_fontain1} at $E=10^{-8}$. (b) Same as in (a) but when
$E=10^{-10}$. In (a) we have superposed the path of characteristic started northward at the equator taking the same frequency of the mode.
}
   \label{ec_mode1}
\end{figure}

\begin{figure} \centering
   \begin{subfigure}{.49\textwidth} \vfill \centering
      \includegraphics[width=\linewidth,clip=true]{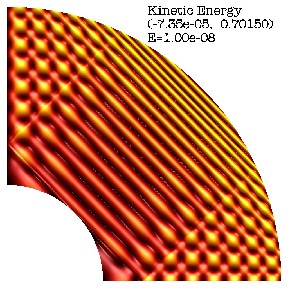}
      \caption[]{}
   \end{subfigure} \hfill
   \begin{subfigure}{.49\textwidth} \vfill \centering
      \includegraphics[width=\linewidth,clip=true]{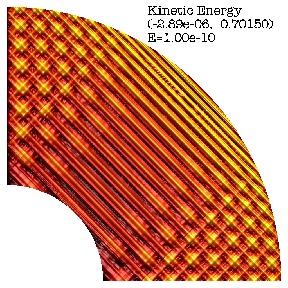}
      \caption[]{}
   \end{subfigure}\hfill
   \caption[]{
Same as figure~\ref{ec_mode1} but for the mode tagged ``6" in figure
~\ref{sp_fontain1}.
}
   \label{ec_mode6}
\end{figure}

First, we may observe that the non-symmetric distribution of
eigenvalues with respect to the line $\omega=\sin(\pi/4)$ actually
reflect an alternate distribution of the branches on each side of the
$\omega=\sin(\pi/4)$-line as shown by Fig.~\ref{repliee}. Second, we observe
that the actual eigenmodes are featured by the web of characteristics.
The modes show periodic  structures (e.g. figures~\ref{ec_mode1} and
\ref{ec_mode6}) even if characteristics do not follow periodic orbits
(this is possible only when  $\omega=\sin(\pi/4)$).

\begin{figure} \centering
   \begin{subfigure}{.48\textwidth} \vfill \centering
      \includegraphics[width=\linewidth,clip=true]{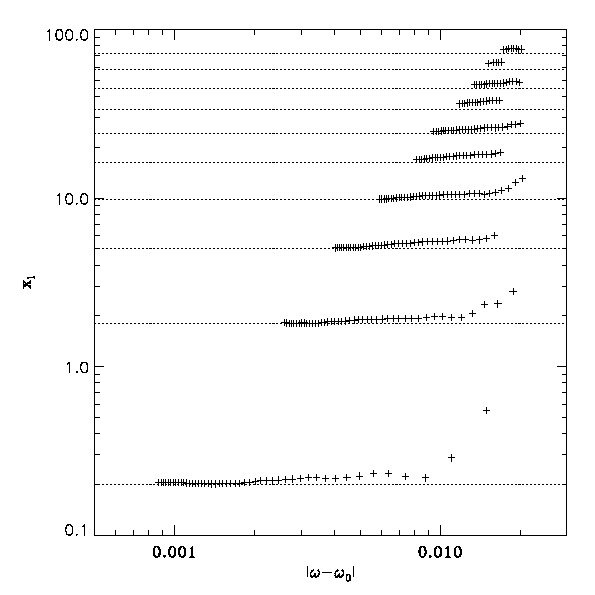}
      \caption[]{}
   \end{subfigure} \hfill
   \begin{subfigure}{.48\textwidth} \vfill \centering
      \includegraphics[width=\linewidth,clip=true]{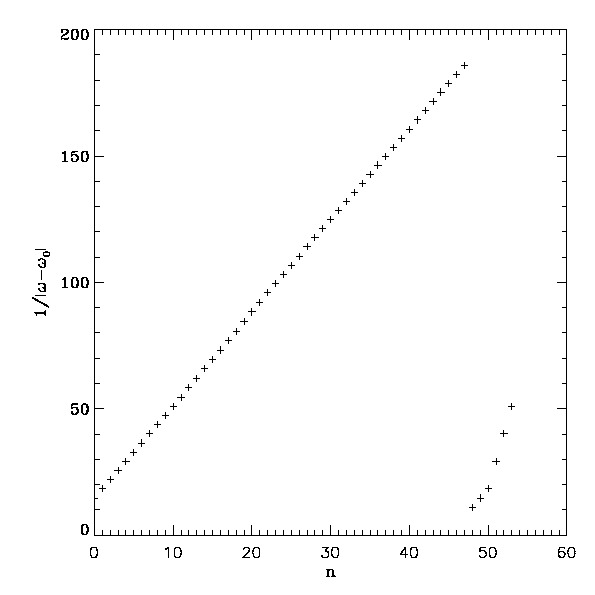}
      \caption[]{}
   \end{subfigure}
   \caption[]{
(a) Quantization of the branches: for all the
eigenvalues shown in Fig.~\ref{sp_fontain1} the quantity
$(\omega-\sin(\pi/4))^2\tau/E$ is plotted versus $|\omega-\sin(\pi/4)|$
(pluses). The dotted lines show the expected formula
$x_\ell=\frac{4}{5}\lp\ell+\demi\rp^2$.
(b) Quantization along the first branch: $1/|\omega-\sin(\pi/4)|$ versus
the order of the eigenvalue.
}
   \label{quantiz}
\end{figure}

The propagation of characteristics associated with a mode frequency is
nevertheless interesting. Characteristics are indeed showing the
path of energy but also the location of equiphase lines (recall
that the group and phase velocities are orthogonal for inertial
waves). Hence, the distance between two parallel characteristics (e.g.
figure~\ref{ec_mode1}a or \ref{ec_mode6}a) may be interpreted as the typical
wavelength of the mode. Let us consider a mode whose frequency is
slightly different from $\sin(\pi/4)$. Let say that

\beq \omega = \sin\lp\frac{\pi}{4}\pm\eps\rp, \with \eps\ll1
\eeqn{omrel}
The distance between the two characteristics of negative slope with one
issued from the equator of the outer sphere is 

\[ \Lambda = \frac{\sin(2\eps)}{\sin(\pi/4)}\]
but from~\eq{omrel} we have

\[ \omega-\sin(\pi/4) = \pm\eps\cos(\pi/4)\]
at first order. Hence, the typical wavelength of the mode is

\beq \Lambda \simeq 4|\omega-\sin(\pi/4)|
\eeqn{B}
where we assumed $\Lambda>0$. Thus the damping rate of the mode should
scale like

\[ \tau \sim -4\pi^2E/\Lambda^2 = \frac{E\pi^2}{4(\omega-\sin(\pi/4))^2}\]
Thus for a given set of modes (a branch) we expect that

\beq \tau(\omega-\sin(\pi/4))^2 = -Ex_\ell\eeq
where $x_\ell$ is a constant specific to the branch. As shown in
figure~\ref{quantiz}a, $x_\ell$ is indeed a constant. Actually, the
constant $x_\ell$ is quantized in a simple way:

\beq x_\ell \simeq \frac{4}{5}(\ell+1/2)^2, \with \ell=0, 1, \ldots
\eeq
This expression is derived from a numerical fit. It betrays again the
quantization of the harmonic oscillator, showing that each branch
corresponds to a different state of this oscillator. Unfortunately, we
could not recover this formula from a simple theoretical argument.

\begin{figure} \centering
   \begin{subfigure}{.48\textwidth} \vfill \centering
      \includegraphics[width=\linewidth,clip=true]{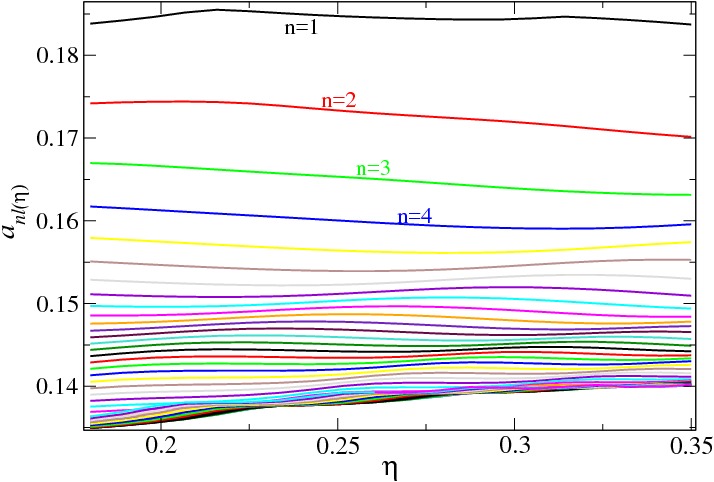}
      \caption[]{}
   \end{subfigure} \hfill
   \begin{subfigure}{.48\textwidth} \vfill \centering
      \includegraphics[width=\linewidth,clip=true]{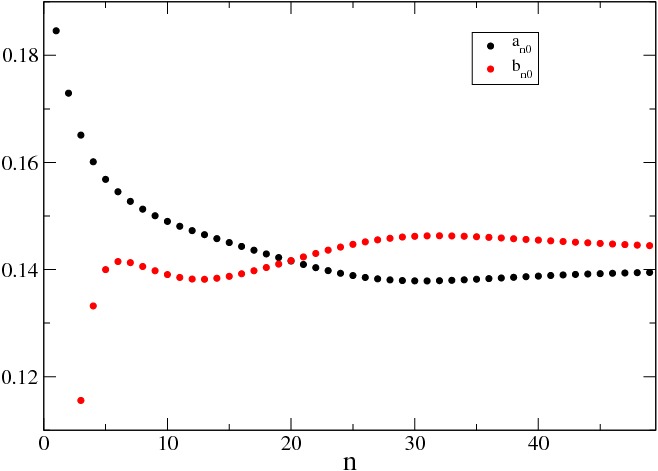}
      \caption[]{}
   \end{subfigure}
   \caption[]{
(a) Variation with $\eta$ of the eigenvalues belonging to
the red branch of Fig.~\ref{sp_fontain1}. The curves plotted are $
a_{n\ell}(\eta)=\frac{n\left[\omega_{n\ell}-
\sin(\pi/4)\right]}{(1-\eta)(\ell+1/2)}$
for $n<50$ and $\ell=0$.  The top curves correspond to the lowest
values of $n$, thus to the least damped eigenvalues.  (b) black:
for each $n$ we plot the best fit of coefficient $a_{n0}(\eta)$ of
(\ref{eq:modes_reguliers_formule_simple}) as obtained from the data
plotted in (a).  Similarly we plot in red the best fit of coefficient
$b_{n0}(\eta)$ appearing in (\ref{eq:modes_reguliers_formule_simple}).
}
   \label{modes_reguliers_formules}
\end{figure}

We may however proceed a little further if we look for the quantization
along a given branch. Comparison of two modes of a branch (e.g.
figure~\ref{ec_mode1} \& \ref{ec_mode6}) shows that they differ by their
typical wavelength along the radius\footnote{For modes in the second
branch (green dots in Fig.~\ref{repliee}a) with a similar frequency as
the modes of the first branch (red dots), the same shape as in
figure~\ref{ec_mode1} is observed but the dominant wavenumber is
increased by some factor.}. Thus, we also should expect from \eq{B} that

\[ |\omega-\sin(\pi/4)| \propto \frac{1-\eta}{n} \; .\]
Fig.~\ref{quantiz}b shows the linear behaviour of $1/|\omega-\sin(\pi/4)|$
with the rank of the eigenvalues of a branch. Hence, eigenvalues on a
given branch seem to verify

\beq
|\omega_{n\ell}-\sin(\pi/4)| =\frac{a_{n\ell}(\eta)(1-\eta)(\ell+1/2)}{n}
\qquad
\tau_{n\ell} = -\frac{En^2}{\left[ b_{n\ell}(\eta)(1-\eta)\right]^2}
\label{eq:modes_reguliers_formule_simple}
\eeq
with $a_{n\ell}(\eta)$ and $b_{n\ell}(\eta)$ real values that have
a very mild dependence on $\eta$.  Their independence on $\eta$ is
checked in Fig.~\ref{modes_reguliers_formules}a: the curves are almost
flat horizontal lines.  From this figure we also see that the curves
for larger $n$ (that are the lower ones) cluster very near the same
value: this means that $a_{n\ell}$ and $b_{n\ell}$ do not vary much
with $n$ (for large enough $n$).  We also checked that they do not
depend much on $\ell$.  This is shown in Fig.~\ref{repliee}a where
we plot the eigenvalues given by the above formula using constant
values  for $a$ and $b$, precisely those computed by best fit of formula
\eq{eq:modes_reguliers_formule_simple} with the actual eigenvalues shown
in Fig.~\ref{sp_fontain1}. The best fit gives $a=0.141$ and $b=0.152$,
and we see from Fig.~\ref{repliee}a that the actual and predicted
eigenvalues match quite well, at least for $n$ large enough.

Actually an even better fitting formula for the spectrum in this  region
is found to be

\beq \tau_{n\ell} = -E(9.89n-0.70\ell+0.26)^2\qquad 
|\omega_{n\ell}-\sin(\pi/4)| = \frac{\ell+1/2}{11.35n-2.33\ell-15.02}
\label{eq:modes_reguliers_formule_compliquee}
\eeq
It reproduces fairly well a large fraction of the eigenvalues as shown in
Fig.~\ref{repliee}b.

In Fig.~\ref{modes_reguliers_formules}b we show the best fit of $a_{n0}$
(black points) and $b_{n0}$ (red points) obtained using the computed
eigenvalues for $\eta$ in the range $0.18\leq \eta\leq 0.35$.  For a
given value of $n$ the fit is done by computing the values $a_{n0}$
and $b_{n0}$ that minimize the error on $a_{n0}-a_{n0}(\eta)$ and
$b_{n0}-b_{n0}(\eta)$.  We confirm that, apart from the first values of
$n$ for which we have remarked previously that the smooth behaviour at
$E=10^{-8}$ is already lost (see Fig.~\ref{sp_fontain1}b), these values
depend very little on $n$, and moreover that they tend to very similar
values for large $n$: for example $a_{49,0}=0.139$ and $b_{49,0}=0.145$.

\begin{figure} \centering
   \begin{subfigure}{.55\textwidth} \vfill \centering
      \includegraphics[width=\linewidth,clip=true]{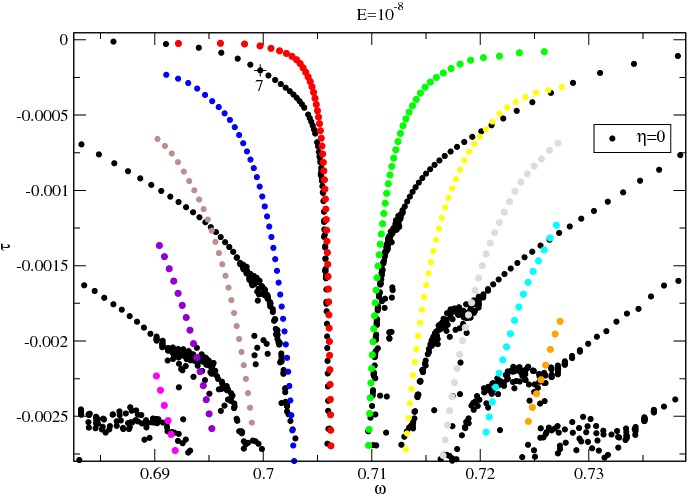}
      \caption[]{}
   \end{subfigure} \hfill
   \begin{subfigure}{.41\textwidth} \vfill \centering
      \includegraphics[width=\linewidth,clip=true]{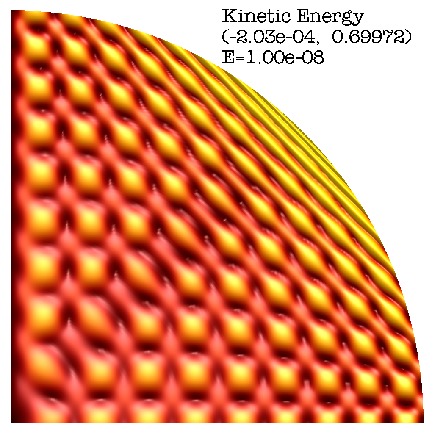}
      \caption[]{}
   \end{subfigure}
   \caption[]{
(a) Spectrum near the frequency $\sin(\pi/4)$ at Ekman number $E=10^{-8}$.
Black dots correspond to the full sphere $\eta=0$. Colored dots are for
the spherical shell with aspect ratio $\eta=0.35$. They are the same
as in Fig.~\ref{sp_fontain1}.  (b) Kinetic energy for mode 7 of first
branch at $\eta=0$, the eigenvalue marked with a plus sign in panel (a).
Note the similarity with the shape of eigenvalue 6 of the red family,
shown in Fig.~\ref{ec_mode6}(a).
}
   \label{ffspect}
\end{figure}

The foregoing results suggest that the eigenvalue spectrum has some
universal features around $\sin(\pi/4)$ independent of the aspect ratio
of the shell. Thus, we examined the eigenvalue spectrum of the full
sphere ($\eta=0$), for which eigenmodes exist even at zero Ekman number.
The result is shown in figure~\ref{ffspect}a where we superpose the
spectra of the spherical shell at $\eta=0.35$ and of the full sphere
($\eta=0$).  The noise in the very damped modes of full sphere are
due to roundoff errors.  Nevertheless, we clearly recognize that the
eigenvalues of the full sphere are also distributed in branches like
those of the spherical shell. Corresponding branches of the full sphere and
the spherical shell tend to merge in the strongly damped part of the spectrum.
Figure~\ref{ffspect}b shows that the shape of a strongly damped regular
mode of the full sphere is very similar to that of a quasi-regular mode
of the shell (e.g. figure~\ref{ec_mode6}).

\begin{figure} \centering
   \begin{subfigure}{.48\textwidth} \vfill \centering
      \includegraphics[width=\linewidth,clip=true]{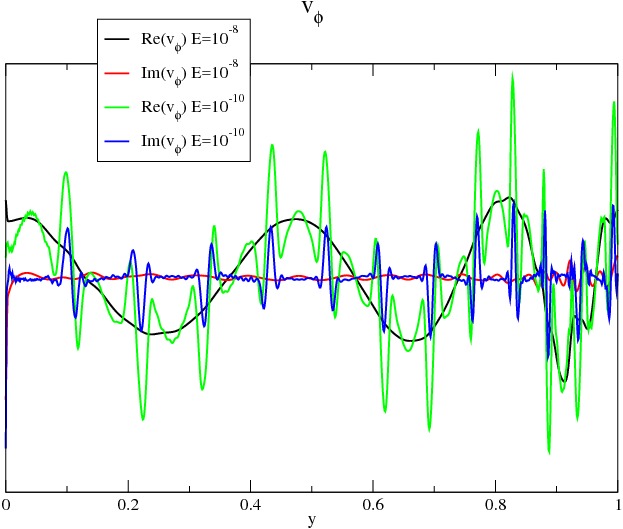}
      \caption[]{}
   \end{subfigure} \hfill
   \begin{subfigure}{.48\textwidth} \vfill \centering
      \includegraphics[width=\linewidth,clip=true]{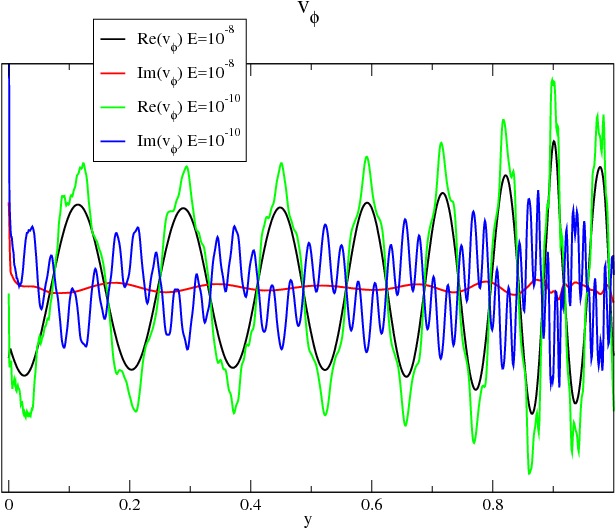}
      \caption[]{}
   \end{subfigure}
   \caption[]{
Profile of $v_\phi$ for mode tagged ``1" in fig.~\ref{sp_fontain1} (a) and mode 6 (b). The profile is shown along the radial direction starting 
from inner critical latitude. Note the smoothness of the black
lines ($E=10^{-8}$) and the emergence of small-scales in the green
lines ($E=10^{-10}$).
}
   \label{vphi_cut}
\end{figure}

\begin{figure} \centering
   \begin{subfigure}{.48\textwidth} \vfill \centering
      \includegraphics[width=\linewidth,clip=true]{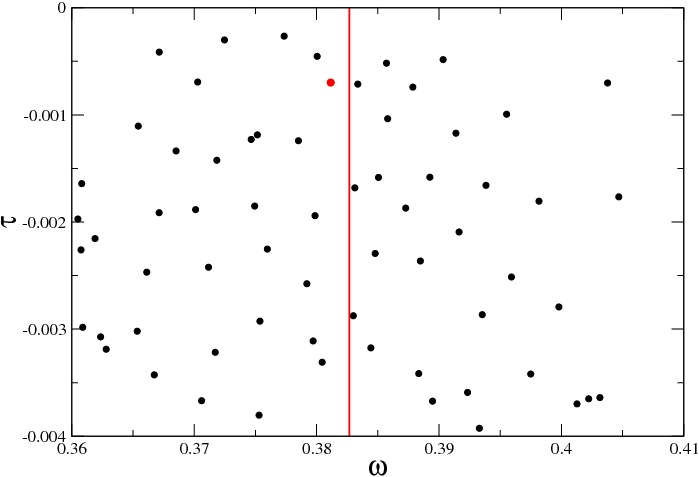}
      \caption[]{}
   \end{subfigure} \hfill
   \begin{subfigure}{.48\textwidth} \vfill \centering
      \includegraphics[width=\linewidth,clip=true]{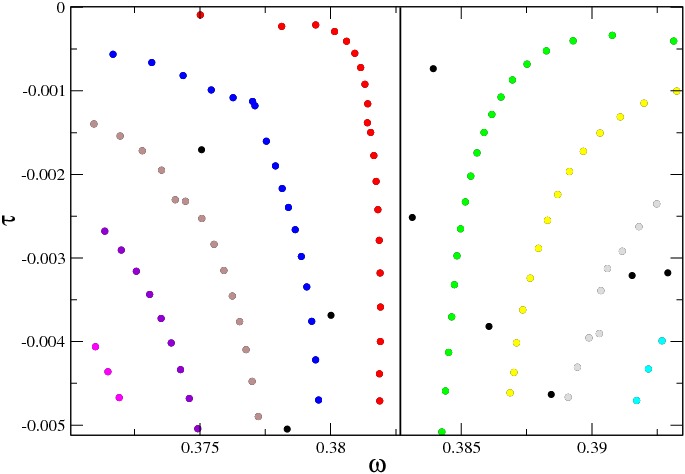}
      \caption[]{}
   \end{subfigure}
   \caption[]{
Eigenvalue spectrum in the complex plane around $\omega=\sin(\pi/8)$
when $\eta=0.35$ (a), and when $\eta=0.2$ (b).
}
   \label{pisur8}
\end{figure}

The foregoing results explain the ``anti-resonance" observed by
\cite{RV10} on tidally forced inertial modes when the forcing frequency
equals $\sin(\pi/4)$. As the forcing frequency tends to $\sin(\pi/4)$
the wavenumber of the excited modes tends to infinity freezing any response
of the fluid to a periodic forcing at $\omega=\sin(\pi/4)$.

Let us now comment on the disappearance of the regular nature of the modes
when viscosity is reduced.  Figures~\ref{ec_mode1}b, \ref{ec_mode6}b
and \ref{vphi_cut} show that the structure of the modes changes below
some $E$ specific to the mode: Small-scale features appear if the Ekman
number is small enough. We interpret this behaviour as follows: the wave
energy propagates along trajectories shown by characteristics. The
lower the Ekman number the longer the wave can propagate without
damping. Close to $\sin(\pi/4)$, characteristics trajectories may be
very long before they hit the inner boundary because the closer $\omega$
to $\sin(\pi/4)$ the longer the trajectory (see Fig.~\ref{ec_mode1}c
and \ref{ec_mode6}c). As long as the wave does not touch the inner
shell, the mapping governed by the characteristics does not change
the scale of the wave \cite[][]{RGV01}.  Hence, for a given value of
$|\omega-\sin(\pi/4)|$, if the Ekman number is large enough, the wave
amplitude has enough time to decrease when it hits the inner shell,
so that the small scales generated by the reflections do not show up
in the mode, which then shows a quasi-regular pattern (black lines
in Fig.~\ref{vphi_cut}). On the contrary, if $E$ is small enough,
propagation along characteristics is almost without attenuation and the
wave hits the inner sphere with nearly its initial amplitude, making
small-scale features generated by the reflections clearly visible and
influential (green curves in Fig.~\ref{vphi_cut}). When we select the
least-damped modes, we select the modes where small scales have the
least amplitude. Hence, for a given $E$, quasi-regular modes only exist
in some neighbourhood of $\sin(\pi/4)$, where characteristic paths not
hitting the inner shell are long enough. We thus deduce that in the
limit $\omega\tv\sin(\pi/4)$ the quasi-regular nature of the modes can
be conserved asymptotically for $E\tv0$ but at the price of considering
modes with higher and higher wavenumbers as imposed by the web of
characteristics (compare Fig.~\ref{ec_mode1}c and \ref{ec_mode6}c).

\begin{figure} \centering
   \begin{subfigure}{.48\textwidth} \vfill \centering
      \includegraphics[width=\linewidth,clip=true]{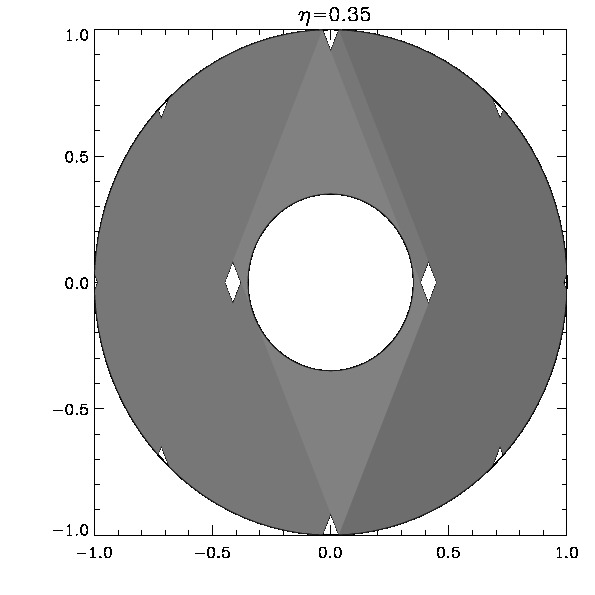}
      \caption[]{}
   \end{subfigure} \hfill
   \begin{subfigure}{.48\textwidth} \vfill \centering
      \includegraphics[width=\linewidth,clip=true]{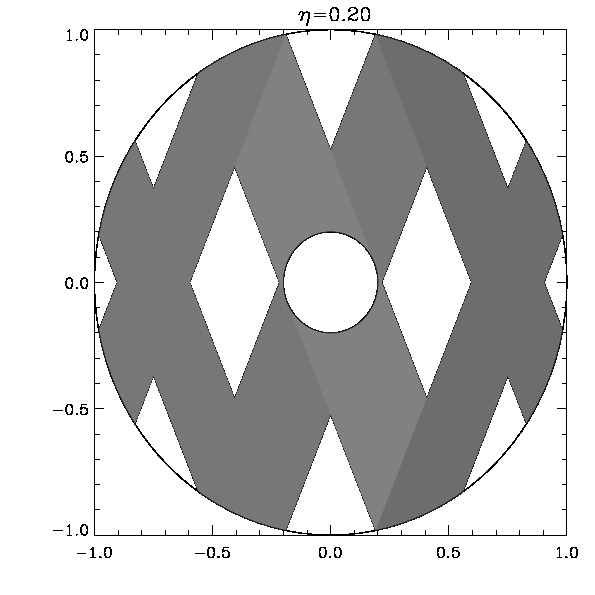}
      \caption[]{}
   \end{subfigure}
   \caption[]{
Shadow of the core when $\omega=\sin(\pi/8)$ for $\eta=0.35$ (a)  and
$\eta=0.2$ (b).
}
   \label{shadows}
\end{figure}

We may now wonder whether the previous results obtained for modes with a
frequency around $\sin(\pi/4)$ extend to other frequencies associated
with periodic orbits. $\sin(\pi/4)$ indeed gives periodic orbit whatever
the radius of the inner shell. As shown in \cite{RGV01}, other
periodic orbits are possible if the radius of the inner core is small
enough. In figure~\ref{pisur8} we show the neat transformation of
the spectrum around $\omega=\sin(\pi/8)$ when the radius of the
inner core is decreased from 0.35 to 0.20. When $\eta=0.35$ periodic
orbits are possible, but mainly in the shadow path of the core (see
Fig.~\ref{shadows}). When the core is smaller, periodic orbits similar
to those of the full sphere (i.e. that never hit the inner boundary)
have a larger phase space that authorize modes with larger scales to
exist, and we recover a spectrum structure that is similar to that of
the full sphere (compare Fig.~\ref{pisur8}b and Fig.~\ref{ffspect}). When
$\eta=0.35$, it is likely that a similar structure exists, but at scales
that are not reachable numerically.

The quasi-regular nature of the modes around frequencies $\sin(p\pi/q)$
that are allowed by the size of the core is however not systematic.
While investigating the case of $\omega=\sin(\pi/6)$, which is associated
with periodic orbits when $\eta\leq0.5$, we found that no regular
quantization occurs if $\eta=0.35$ but also if $\eta=0.20$. Inspection of
the eigenmodes shows that the critical latitude singularity on the inner
sphere is excited, hence inserting new scales in the eigenfunctions and
precluding any regular behaviour as well as simple quantization rules.
The reason why this occurs for this periodic orbit and not the others is
not clear.

\section{Analysis of attractor modes}

We now analyse the dynamics of the flows that structures the
shear layers looping around the attractors.

\begin{figure} \centering
\includegraphics[width=0.88\linewidth,clip=true]{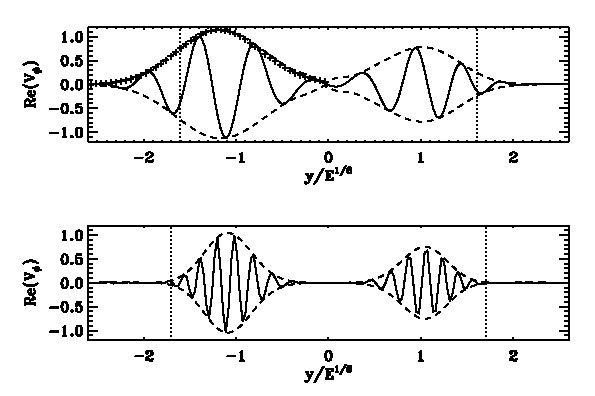}
\caption[]{Cut through the shear layers of the least-damped mode
associated with the attractor displayed in Fig.~\ref{m782}. Top:
$E=10^{-8}$. Bottom: $E=10^{-11}$ The origin of the $y$-coordinate is
the position of the asymptotic attractor while the dotted vertical lines
show the position of the actual attractor. The dashed curves show the
envelope of the wave packet. On the top figure the pluses show the
rescaled envelope of the $E=10^{-11}$-solution assuming an $E^{1/4}$-scaling
law for the width of the envelope.}
\label{coupe782}
\end{figure}

\subsection{Summary of numerical results}

\begin{figure} \centering
\includegraphics[width=0.7\linewidth,clip=true]{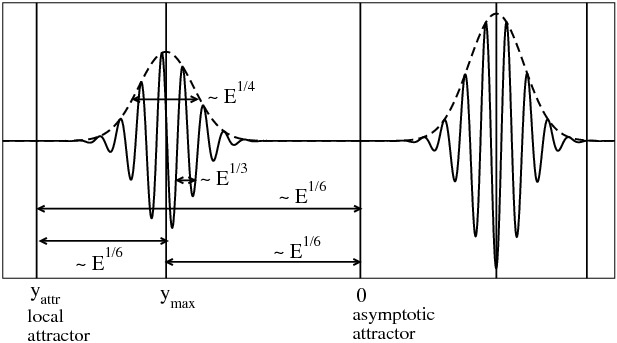}
\caption[]{Schematic view of the cut through shear layers looping around
the attractors, showing up the three scales involved: $E^{1/6}$ is the
scale governing the distance among local attractor, asymptotic attractor
and the shear layer.  $E^{1/4}$ gives the extension of the shear layer.
$E^{1/3}$ is the wavelength of the oscillations inside the shear layer.
Note that both the local attractor and the asymptotic attractor are
positioned (asymptotically as $E\to 0$) outside the shear layer.
}
\label{figure_schematique}
\end{figure}

The first hint given by the numerical solutions is the law \eq{egv_exp}
governing the eigenvalues associated with attractor modes. From this law
we note that the frequency shift of the modes with respect to the
asymptotic frequency $\omega_0$ is
$\Im(\lambda-i\omega_0)=\od{E^{1/3}}$. Since the distance between the
actual attractor and the asymptotic one varies like
$\sqrt{|\omega-\omega_0|}$ \cite[][]{RGV01}, we deduce that this
distance scales like $E^{1/6}$. Besides, as already shown by
\cite{K95}, the scale $E^{1/3}$ turns out to be the smallest scale of
detached shear layers. However, \cite{RV97} also noticed that
some shear layers display a thickness scaling with $E^{1/4}$. We can
illustrate the presence of these three scales using modes associated
with the asymptotic attractor whose frequency is $\omega_0=0.782$
(when $\eta=0.35$). Figure~\ref{m782} (left) displays the
shape of the least-damped mode associated with this attractor.
In figure~\ref{coupe782} we show the variations of the amplitude of the
velocity component $u_\varphi$ in the transverse direction of the shear
layer. The three scales are clearly showing up. Indeed, taking the
origin of the coordinate at the asymptotic attractor, and rescaling the
coordinate with
$E^{1/6}$, we see that the position of the maximum does not change when
the Ekman number drops from $10^{-8}$ to $10^{-11}$. In the top figure,
we note that the rescaled envelope of the $E=10^{-11}$-solution
perfectly matches the $E=10^{-8}$-envelope, hence giving evidence that the
$E^{1/4}$ scale indeed determines the width of the wave packet. Finally,
the ratio of the wavelength, rescaled by $E^{1/6}$, is $\simeq3.12$, not far
from the expected $\sqrt{10}\simeq3.16$ if the wavelength scales as
$E^{1/3}$. Since the three scales also appear as such in other attractor
modes, we shall assume that they are the scales controlling the dynamics
of the shear layers associated with (at least some) attractor modes.
Figure~\ref{figure_schematique} schematically depicts the situation.

\subsection{The reduced problem}

To begin with we recall that system \eq{eqmo} can be cast into a single
equation for the pressure perturbation, namely

\beq
(\lambda - E \Delta)^2 \Delta P + \ddpptwo{P}{z} = 0
\eeqn{poincvisc}
where we assumed solutions of the form

\[ P(\vr,t) = P(\vr)e^{\lambda t}\]
and where $\lambda=\tau+i\omega$ is the complex frequency. $\tau$ is
the damping rate and $\omega$ the real frequency of the mode. Equation
\eq{poincvisc} reduces to Poincar\'e equation when the Ekman number $E$
is set to zero.

As observed by \cite{RV97}, shear layers built on attractors own a
(inviscid) singularity on the rotation axis generating a divergence in
$s^{-1/2}$, where $s$ is the radial cylindrical coordinate. We remove
this divergence by setting $P=ps^{-1/2}$. Thus doing, we rewrite
\eq{poincvisc} as

\beq
(\lambda - E \Delta')^2 \Delta' p + \ddpptwo{p}{z} = 0
\eeqn{poincviscb}
with

\[ \Delta' = \dds{} + \ddz{} + \frac{1}{4s^2}\]
We also use coordinates parallel $x$ and perpendicular $y$ to the shear
layer/attractor branch such that

\beq x=\alpha z +\omega s, \qquad y=\omega z - \alpha s,
\eeq
with $\alpha=\sqrt{1-\omega^2}$. It implies that

\[ \ds{} = \omega\dx{} - \alpha\dy{}, \qquad \dz{} =
\alpha\dx{}+\omega\dy{}, \qquad \ddz{} = \alpha^2\ddx{}
+\omega^2\ddy{} + 2\alpha\omega\dxdy{}  \]
From the numerical solutions we find that (see \eq{egv_exp})

\beq \lambda = i\omega_0+\lambda_1E^{1/3} +
\lambda_2E^{1/2}+\cdots\eeqn{vp_exp}
The numerical solutions have also shown the importance of three
non-dimensional scales: $E^{1/3}$, $E^{1/4}$ and $E^{1/6}$. If we select
the smallest scale $E^{1/3}$ and consider the dominating terms, the full
equation \eq{poincviscb} can be reduced to a simpler equation, which
reads:

\beq
\alpha_0\dx{p} = iE\dddy{p} - i\tau \dy{p}
\eeqn{eq:reduced}
and which we shall call the reduced problem. Its derivation is given
in appendix A. We now analyse this new and simpler but still very rich
equation.  We remark that this equation is the same as the one we obtained
for the slender torus considered in \cite{RVG02}.  The geometry in the
meridional section is also the same. The only difference between the two
cases is that in the spherical geometry the reflection on the rotation
axis produces a variation in the solution: We show in appendix that if
viscosity is neglected each reflection produces a factor $-i$ in the
eigenfunction. This effect is not present in the toroidal configuration,
as there is no axial singularity there. Viscosity may actually alter the
phase shift due to axis-reflection, but we leave this possible effect to
future investigations. Finally, we may note that attractors bouncing
K-times impose a factor $(-i)^K$ in the solution, but this factor
reduces to unity if $K=4n$. As we shall see below, the analytical
2D solutions obtained for the slender torus can offer a very good
approximation to the eigenmodes made of shear layers bouncing $4n$
times on the rotation axis, even in a thick shell.

\subsection{Local dynamics of shear layers}

Let us first assume that the shear layers are formed by the product of
a fast oscillating wave of wavenumber scaling as $\calo{E^{-1/3}}$ and
a wide $\calo{E^{1/4}}$ envelope.  It is convenient to set $p=e^\psi$
and to work with $\psi$.  Equation for $\psi$ is:

\beq
\alpha_0 \dx\psi=iE\left[\ddppthree{\psi}{y}+ 
3\ddpptwo{\psi}{y}\ddpp{\psi}{y}+\lp\ddpp{\psi}{y}\rp^3\right] 
+ iE^{1/3}\abstauhat \ddpp{\psi}{y}
\eeqn{eq:reduced_v_bis}
where we have set $\abstauhat= -\tau \emoinsuntiers$, hence 
$\abstauhat \sim \calo{E^0}$.
We remark that an exact solution to this equation is the linear function:

\beq
\psi(x,y)=\frac{q^3-\abstauhat q}{\alpha_0}x+E^{-1/3}iqy
\eeqn{eq:vone}
where $q$ is any complex constant.  We shall make the assumption that
$q$ is real, so that (\ref{eq:vone}) describes a wave with  spatial
frequency $\frac{q \emoinsuntiers}{2 \pi}$ in the $y$ direction and
exponential variation in the $x$ direction.  This function accounts for
the $\euntiers$ oscillations of the shear layers observed numerically.
The foregoing solution is not localized and we therefore need to seek
for the envelope of the wave that keeps it close to the attractor.
Recalling that $p=e^\psi$, we now set

\beq
\psi=\frac{q^3-\abstauhat q}{\alpha_0}x+\emoinsuntiers iqy+h(x,y)
\eeqn{psi}
and we shall assume that $h\sim\calo{E^0}$,
$\frac{\partial^n h}{\partial y^n}\sim E^{-n/4}$, $\dx{h}\sim E^{1/12}$.
We insert this expression into (\ref{eq:reduced_v_bis}).
To leading order, that is $\calo{E^{1/12}}$,
we get the following equation for $h$:
\[
\alpha_0\dx{h}=iE^{1/3}\big(\abstauhat-3q^2\big)\dy{h}
\]
This first order linear partial differential equation 
has the following general solution:

\beq
h(x,y)=f(Z),\quad {\rm with}\quad Z=\frac{i(\abstauhat-3q^2)}{\alpha_0}E^{1/12}x+E^{-1/4}y
\label{eq:vtwo}
\eeq
where $f(Z)$ is an arbitrary function.
Numerical solutions suggest that the envelope is a Gaussian, which would be 
the case if $f(Z)=-aZ^2$ for some complex coefficient $a$ with a positive 
real part. We 
shall prove in the following that $f(Z)$ has indeed this shape and 
we shall provide the expression of $a$ as a function of the  eigenvalue and of the 
geometrical parameters 
of the attractor (see Eq.~\ref{hz}).

\subsubsection{Viscous evolution along a characteristic path}

We now wish to obtain the variation of the solution after travelling one
complete loop along the attractor.
After reflection on a boundary, $p$ is rescaled as

\beq
p_n(x_n,y)=K_n p\left(x_1,\frac{y}{K_n}\right)
\label{eq:ricoche_rescaling}
\eeq
where $K_n$ are products of contraction/dilation coefficients
arising from the reflections on the boundaries. 
They are the same as
those of \cite{RVG02}\footnote{Note that there is a missprint in page 354 of
\cite{RVG02}: the third formula of that page should read like
(\ref{eq:ricoche_rescaling}) instead of having $K_n$ at the
denominator twice.}. Therefore the spatial frequency $q$ is multiplied at
each reflection by the factor $1/K_n$.
Taking the first branch as the reference, we set $K_1=1$. 
The variation of the $E^{1/3}$ part of the solution (\ref{eq:vone})
after travelling over the length $\delta x=\ell_n$ on the branch $n$ is
exactly:

\[\delta \psi_n=\frac{(q/K_n)^3-\abstauhat q/K_n}{\alpha_0}\ell_n\]
Since $h$ varies by a small \od{E^{1/12}} amount over each branch,
we can use the same procedure as in section (3.2) of \cite{RVG02}
to determine its variation after one loop. On branch $n$ we have:

\beqan
\delta h_n &=& h_n(x_n+\ell_n,y)-h_n(x_n,y)
\simeq
\left.\ddpp{h}{x}\right|_{x_n,y} \ell_n
\\
&=& \frac{iE^{1/3}}{\alpha_0}\left(\abstauhat-\frac{3q^2}{K_n^2}\right) 
   \ddpp{h(x_1,y/K_n)}{y}
\ell_n
\eeqan{horvar}
Since the functions $h_n$ are the same on every branch up to a scale factor
$K_n$, we can write \eq{horvar} as

\beqa
\delta h_n &=&
\frac{iE^{1/3}}{\alpha_0}\left(
   \frac{\abstauhat}{K_n}-\frac{3q^2}{K^3_n}\right)\ddpp{h(x_1,y/K_n)}{(y/K_n)} \ell_n \\
&=& \frac{iE^{1/3}}{\alpha_0}\left(
\frac{\abstauhat}{K_n}-\frac{3q^2}{K^3_n}\right)
   \ddpp{h(x_1,y)}{y}\ell_n 
\eeqa
where all the derivatives are taken on the first branch.


Finally we have to take into account the variations due to the
reflections on the rotation axis: as we
show in appendix, each reflection
 introduces a factor $-i$ in the eigenfunction;
therefore $\psi$ is shifted by $-i \pi/2$ at each reflection. 

Summing up all the contributions of the
perturbations arising from all the branches of the attractor we get:

\beq
\delta \psi^{\rm propag.} =
\underbrace{
-\frac{iK\pi}{2}+\frac{Aq^3-B\abstauhat q}{\alpha_0}}_{\calo{E^0}}+
\underbrace{
\frac{iE^{1/3}}{\alpha_0}\left(B\abstauhat-3Aq^2\right)\dy{h}}_{\calo{E^{1/12}}}
\eeqn{viscvar}
where $K$ is the total number of reflections on the axis, and

\[ A=\sum_n{\frac{\ell_n}{K_n^3}}, \qquad B=\sum_n{\frac{\ell_n}{K_n}}\;
.\]
We remark that the values of $A$ and $B$ depend on the branch 
of the attractor that is chosen as the first branch.
The quantity $d=B^3/A$ however does not change.
Expression \eq{viscvar} gives, up to terms of order $E^{1/12}$, the
variation of the perturbation due to viscosity when the wave owns the
$E^{1/3}$ and $E^{1/4}$ scales. Except for the reflections on the
boundaries, which rescale the width of the layer by some factor of order
unity, the foregoing expression is just an approximate solution of
\eq{eq:reduced}. We now need to take into account the fact that the
perturbation is not strictly on the attractor and therefore that after
one loop the place where we measure the variation $\delta \psi$ is not the
same as the initial one: it has been shifted by a small amount
controlled by the mapping. Indeed, the only point 
that comes back to the same place is the one on the attractor.

\subsection{The part played by the mapping}

To take into account the shift induced by the mapping, we use the same
procedure as the one devised in \cite{RVG02}. Indeed, the mapping drawn
by characteristics in the meridional plane of the spherical shell is the
same as the one of the slender torus used in \cite{RVG02}.

Here too, we shall work with the associated critical latitude
$\theta^c=\arcsin\omega$, rather than with the frequency $\omega$. Thus,
$\theta^c_0$ designates the critical latitude associated with the
frequency $\omega_0$ of the asymptotic attractor.

We introduce the mapping as the function $f(\phi,\theta^c)$
that associates the latitude $\phi$ where the characteristic
bounces on the inner or outer boundary to the latitude of its image
after one loop along the attractor. The $y$-coordinate
introduced in \eq{eq:reduced} is related to the latitude $\phi$ by

\[ y=p(\phi-\phi_0) \with p=r \sin(\phi_0\pm \theta^c_0)\]
where $\phi_0$ is the latitude of the reflection point of the asymptotic
attractor and $r$ is the radius of the reflecting sphere (either $\eta$ or
$1$). The $\pm$ sign denotes the sign of the slope of the chosen
characteristic. Finally, we note that $\phi_0$ is also the
fixed point of the mapping when  $\theta^c=\theta^c_0$.

Since the mapping just displaces the points, 
its action on
the velocity field complies with

\beq
u(f(y,\theta^c),\theta^c)df = u^{\text{propag.}}(y,\theta^c) dy
\eeqn{umap}
where $ u^{\text{propag.}}(y,\theta^c)$ is the flow field
obtained after propagation along the map with starting point $(x_1,y)$.
Here and in the following $x_1$ will be omitted. $u$ can be understood
as the toroidal component of velocity multiplied by the square root of
the distance to the rotation axis $s^{1/2}$. It admits the same evolution
equation (\ref{eq:reduced}) as the reduced pressure and thus has the
same solution.

In order to find the displacement due to the mapping
we make a Taylor expansion of the mapping around the fixed point of
the asymptotic attractor, namely around $\phi=\phi_0$ (that is $y=0$)  and
$\theta^c=\theta^c_0$. Following the appendix of \cite{RGV01}, we get:

\beq f(\phi,\theta^c) = \phi + f_{01}\delta\theta+\demi
f_{20}(\phi-\phi_0)^2 + \demi f_{02}\delta\theta^2 +
f_{11}\delta\theta(\phi-\phi_0)+ \cdots\eeq
where $\delta\theta=\theta^c-\theta^c_0$. We defined

\[
f_{ij}\equiv
   \left.\frac{\partial
^{i+j}f}{\partial\phi^i\partial\theta^j}\right|_{\phi_0,\theta_0}
\]
We recall that numerical solutions say that $\delta\theta = \calo{E^{1/3}}$
and $\phi-\phi_0 = \calo{E^{1/6}}$.
The foregoing expression of $f$ leads to

\beq
f(y,\theta) = p(f-\phi_0) = y + 
pf_{01}\delta\theta+\frac{f_{20}}{2p}y^2 +\cdots
\eeqn{hf}

It is convenient to shift the $y$ coordinate and develop around 
$\newy=y-\ymax$,
where $\ymax$ is the position where the wave packet amplitude is maximum
(see figure \ref{figure_schematique}). 
We shall assume  $\newy\sim\od{E^{1/4}}$ and we shall drop all the contributions
 smaller than $\calo{E^{5/12}}$.
We thus have:

\beq
f(y,\lambda^c)= \underbrace{\ymax}_{E^{2/12}}
+\underbrace{\newy}_{E^{3/12}}
+\underbrace{pf_{01}\delta\theta+\frac{f_{20}}{2p}\ymax^2}_{E^{4/12}}
+\underbrace{\frac{f_{20}}{p}\ymax\newy}_{E^{5/12}}+\cdots
\eeqn{newf}
We rewrite \eq{umap} in terms of the exponent $\psi$:

\beq
e^{\psi(f(y,\theta^c),\theta^c)}df = e^{\psi^{\text{propag.}}(y,\theta^c)} dy
\eeqn{mapping_phi}
We remark that 

\[
\left.\frac{df}{dy}\right\vert_{\newy=0}=1+\frac{f_{20}}{p}\ymax+\cdots=
e^{\frac{f_{20}}{p}\ymax+\cdots}
\] 
where $f_{20}\ymax/p$ is of order $E^{1/6}$.
Therefore (\ref{mapping_phi}) is turned simply into:

\beq
\psi(f(y,\theta^c),\theta^c) = \psi^{\text{propag.}}(y,\theta^c)-\frac{f_{20}}{p}\ymax+\cdots
\eeqn{mapping_phi_2}
We evaluate separately the l.h.s. and the r.h.s. of (\ref{mapping_phi_2}).
For the l.h.s. we replace $\psi$ with \eq{psi}:

\beqa
\psi(f(y,\theta^c),\theta^c)=&&
\frac{q^3-\abstauhat q}{\alpha_0}x_1+
E^{-1/3}iq(y+pf_{01}\delta\theta+\frac{f_{20}}{2p}\ymax^2+\frac{f_{20}}{p}\ymax\newy)+
\\&&+h(x_1,y)+
\ddpphy\big(pf_{01}\delta\theta+\frac{f_{20}}{2p}\ymax^2+\frac{f_{20}}{p}\ymax\newy)+\cdots
\eeqa
For the evaluation of the r.h.s. we use \eq{viscvar}:

\beqa
&& \psi^{\text{propag.}}(y,\theta^c)=
\psi(y,\theta^c)+\delta\psi^{\text{propag.}}(y,\theta^c)
=
\frac{q^3-\abstauhat q}{\alpha_0}x_1+
\\ &&
+E^{-1/3}iq(\ymax+\newy)+h(x_1,y)-
\frac{iK\pi}{2}\!+\!\frac{Aq^3-B\abstauhat q}{\alpha_0}\!+\!
\frac{iE^{1/3}}{\alpha_0}\left(B\abstauhat-3Aq^2\right)\ddpphy
\eeqa
We are ready to insert these expressions into (\ref{mapping_phi_2});
dropping all the contributions smaller than $E^{1/12}$ we get:

\beqa
&&
\underbrace{E^{-1/3}iq(pf_{01}\delta\theta+\frac{f_{20}}{2p}\ymax^2)}_{E^0}
+
\underbrace{E^{-1/3}iq\frac{f_{20}}{p}\ymax\newy+\ddpphy\big(pf_{01}\delta\theta+\frac{f_{20}}{2p}\ymax^2\big)}_{E^{1/12}}=\\
&&
 -\underbrace{\frac{iK\pi}{2}+\frac{Aq^3-B\abstauhat q}{\alpha_0}}_{E^0}+
\underbrace{
\frac{iE^{1/3}}{\alpha_0}\left(B\abstauhat-3Aq^2\right)\ddpphy
}_{E^{1/12}}
\eeqa
We remark that the term
$\frac{f_{20}}{p}\ymax$
arising in \eq{mapping_phi_2}, due 
to the contraction of the
mapping, is negligible since it is of order $\calo{E^{2/12}}$. 
This equality must be satisfied independently for
the $E^0$ and $E^{1/12}$ terms.  At the lowest order $E^0$ we thus obtain:

\[
-\frac{iK\pi}{2}+\frac{Aq^3-B\abstauhat q}{\alpha_0}
-E^{-1/3}iq\lp pf_{01}\delta\theta+\frac{f_{20}}{2p}\ymax^2\rp=0
\]
and to next order $E^{1/12}$:
\beq
E^{-1/3}iq\frac{f_{20}}{p}\ymax\newy+
\ddpphy\left[pf_{01}\delta\theta+\frac{f_{20}}{2p}\ymax^2-\frac{iE^{1/3}}{\alpha_0}\left(B\abstauhat-3Aq^2\right)\right]=0
\eeqn{disprelE1/12}
Taking the real part of the first relation we find 

\beq \abstauhat = q^2\frac{A}{B}\eeqn{tau-q}
which shows, as expected, that the damping rate is controlled by the
wavelength of the mode.  The imaginary part of the first relation fixes
the position of $\ymax$ in terms of the eigenfrequency and of the geometry
of the attractor:

\beq 
\ymax^2=-\lp\frac{K\pi}{2q}E^{1/3}+pf_{01}\delta\theta\rp
\frac{2p}{f_{20}}
\eeq

The second relation \eq{disprelE1/12} provides the form of $h(y)$
which simply reads

\beq h(y)= \demi a E^{-1/2} (y-\ymax)^2 + b\eeqn{hz}
with

 \beq a = -\frac{q^2\ymax E^{-1/6} f_{20}/p}{iK\pi/2 + 2Aq^3/\alpha_0}\eeq

The shape of $h$ confirms that the wave packet is
localized and with a gaussian shape, as suggested by the numerical solutions.
The gaussian shape is governed by the real part of $a$.
%
Using (\ref{eq:vone}) and (\ref{eq:vtwo}) we finally write the shear
layer profile:

\begin{eqnarray}
&& \label{eq:u_theorique_revisited} u(x,y)= \\
&&u_0\exp \!\la
\frac{(q^3\!-\!\abstauhat q)x}{\alpha_0}\!+\!
iq E^{-1/3}y
 \!+\!\frac{a}{2}\!\left(\!\frac{i(\abstauhat\!-\!3q^2)}{\alpha_0}
 E^{1/12}x\!+\!E^{-1/4}(y-\ymax)\!\right)^{\!\!2}\ra  \nonumber
\end{eqnarray}
where $u_0$ is an arbitrary constant.

\begin{table}
\begin{center}
\begin{tabular}{ccccccccc}
$\omega_0$ & $\alpha_0$  & $c$ & $d$ & $K$ & $\eta$\\
0.555369 & 0.831694 & 43.8 & 88.8 & 2 & 0.35\\
0.831694 & 0.555369 & 43.8 & 88.8 & 2 & 0.35\\
0.622759 & 0.782413 & 49.3 & 28.5 & 2 & 0.35\\
0.782413 & 0.622759 & 49.3 & 28.5 & 2 & 0.35\\
0.466418 & 0.884564 &332.8 & 58.1 & 2 & 0.50\\
0.884564 & 0.466418 &332.8 & 58.1 & 2 & 0.50\\
0.662485 & 0.749075 & 106.3 & 90.1 & 4 & 0.35\\
0.749075 & 0.662485 & 106.3 & 90.1 & 4 & 0.35\\
\hline
\end{tabular}
\caption[]{Geometric parameters for some attractors.}
\label{cdK}
\end{center}
\end{table}

Let us now characterize the position of the local attractor $\yattr$
(see figure \ref{figure_schematique}).
Since on the local attractor we must have
$f(\yattr,\theta^c)=\yattr$, from \eq{hf} we get

\[pf_{01}\delta\theta+\frac{f_{20}}{2p}\yattr^2=0,\]
and thus

\beq
\yattr^2=-\frac{2p^2f_{01}\delta\theta}{f_{20}}
\eeq

The above quantities $\yattr$, $\ymax$, $q$ and $a$ change if the
starting branch along the attractor is changed, because reflections on the
boundaries induce contractions/dilations that are branch-dependent. Hence,
the geometric parameters of the attractor $A$, $B$, $p$, $f_{01}$ and
$f_{20}$ are starting-branch dependent.  However, the two parameters

\beq
c=\left\vert\frac{f_{20}B}{p}\right\vert,\qquad
d=\frac{B^3}{A}
\eeq
are starting-branch independent. Using $|pf_{01}|=B$
\cite[see][]{RGV01} and defining $\hat{\omega}_1\equiv
(\omega-\omega_0)E^{-1/3}=\alpha_0\delta\theta E^{-1/3}$, we can rewrite
the above formulas as follows:

\begin{subequations}

\beq
\frac{\yattr}{ E^{1/6}B}
=\sqrt{\frac{2\absomega1 }{\alpha_0 c}}
\eeq
 
\beq
\left(\frac{\ymax}{E^{1/6}B}\right)^2=
\frac{1}{c}
\left(
\frac{2\absomega1}{\alpha_0}-
\frac{K\pi}{\abstauhat^{1/2} d^{1/2}}
\right)
=
\left(\frac{\yattr}{E^{1/6}B}\right)^2-
\frac{K\pi}{c\abstauhat^{1/2}d^{1/2}}
\eeq

\beq
\frac{\ymax}{\yattr}=
\sqrt{1-\frac{K\pi\alpha_0}{2d^{1/2}\absomega1\abstauhat^{1/2}}}
\label{ymax/yattr}
\eeq

\beq
B^2 a =
-\frac{2\abstauhat \alpha_0 c d |\ymax|/(E^{1/6}B)}
{iK\pi\alpha_0 + 4\abstauhat^{3/2}d^{1/2}}
\eeq
\beq
B^2 \Re(a)=-\frac{8\abstauhat^{5/2}\alpha_0 c d^{3/2} |\ymax|/(E^{1/6} B)}
{K^2\pi^2\alpha_0^2+16\abstauhat^3 d}
\eeq
\beq
q^2 B^2 = \abstauhat d
\eeq
\end{subequations}
In these formulas $B$ is the only geometric parameter that changes when
we change the starting branch and all the quantities on the r.h.s. of
these formula are starting-branch independent.  We remark in particular
that the ratio $\ymax/\yattr$ is independent of the starting branch.
Table~\ref{cdK} gives the starting-branch-independent parameters of the
attractors listed in table~\ref{am}. Finally, note that parameters
$\abstauhat$ and $\absomega1$ have to be given by the numerical
solution.

\begin{figure} \centering
   \begin{subfigure}{.48\textwidth} \vfill \centering
      \includegraphics[width=\linewidth,clip=true]{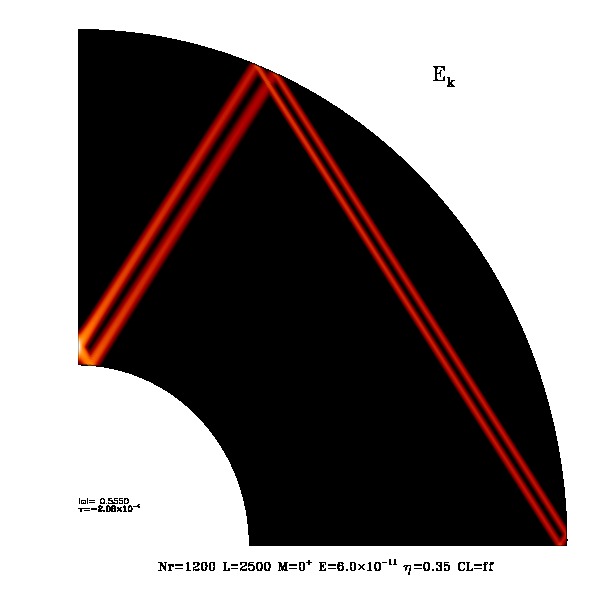}
      \caption[]{}
   \end{subfigure} \hfill
   \begin{subfigure}{.48\textwidth} \vfill \centering
      \includegraphics[width=\linewidth,clip=true]{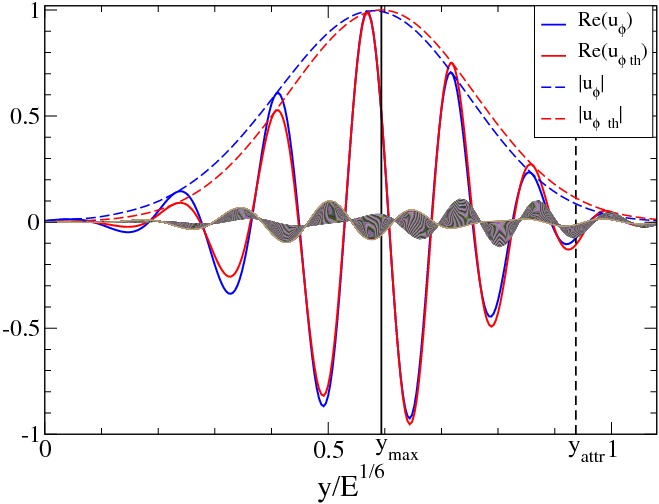}
      \caption[]{}
   \end{subfigure}
\caption[]{(a) Meridional distribution of the kinetic energy of the
least-damped mode associated with the attractor at $\omega_0=0.555369$
when $E=6 \times 10^{-11}$. The green lines show the actual position
of the attractor, while the short and long white lines show the
position of tranverse and longitudinal profiles displayed in (b) and in
Fig.~\ref{u_theo_prof2}a. The dashed rectangle delineate the region where
the difference between the theoretical and the computed solution has
been evaluated. (b) Profile of $Re(u_\phi)$ and $|u_\phi|$ for the same
mode as in (a) together with the profile of the theoretical prediction
(\ref{eq:u_theorique_revisited}) and the difference between them for 345
segments taken inside the rectangle of panel (a). The solid vertical line
visualizes $y_{\rm max}$ and the vertical dashed line shows the position
of the attractor associated with the mode frequency.  The position of
the profile is given by the short white line in (a).
}
\label{u_theo_prof1}
\end{figure}

\begin{figure} \centering
   \begin{subfigure}{.48\textwidth} \vfill \centering
\includegraphics[width=\linewidth,clip=true]{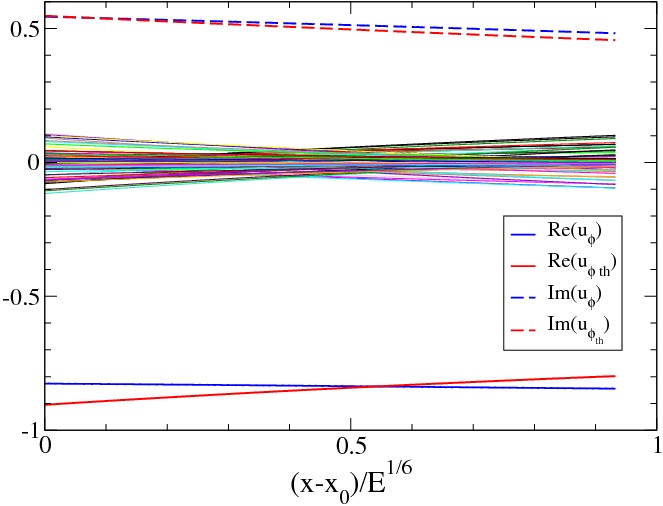}
      \caption[]{}
   \end{subfigure}
   \begin{subfigure}{.48\textwidth} \vfill \centering
\includegraphics[width=\linewidth,clip=true]{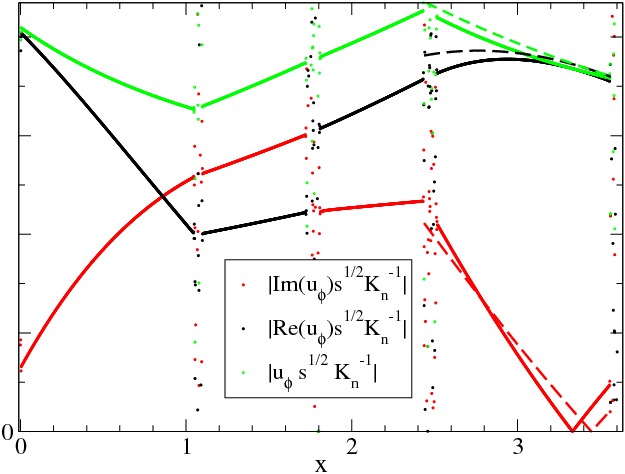}
      \caption[]{}
   \end{subfigure}
\caption[]{(a) Real and imaginary parts of $u_\phi$ from the numerical
and analytical solutions (red and blue lines). The multicolor lines
in the middle show the difference between the theoretical prediction
(\ref{eq:u_theorique_revisited}) and the numerical solution at
various $y$-positions. (b) $u_\phi\sqrt{s}/K_n$ for the mode
shown in Fig.~\ref{u_theo_prof1}a along the associated attractor
(green line in Fig.~\ref{u_theo_prof1}a); $s$ is the distance to
axis and $K_n$ is the amplification factor at the $n^{th}$ bounce
on boundary.  In black is shown the real part, in red the imaginary
part and in green the modulus. The dashed lines in the last interval
(long white segment in Fig.~\ref{u_theo_prof1}a) show the prediction
of~\eq{eq:u_theorique_revisited}.

}
\label{u_theo_prof2}
\end{figure}

\section{Comparison between analytic and numerical solutions}

\subsection{General attractor modes}

In figures \ref{u_theo_prof1} and \ref{u_theo_prof2} we display the
actual eigenfunction for the least-damped eigenmode of attractor
$\omega_0=0.555369$ at $E=6\times 10^{-11}$ and the profile of
the $u_\varphi$ component across (Fig.~\ref{u_theo_prof1}b)
and along (Fig.~\ref{u_theo_prof2}a) the shear layer as given
by (\ref{eq:u_theorique_revisited}) and the numerical solution.
We note the good agreement between the curves: indeed, we expect the
relative difference to be of order of $E^{1/12}$, which is 0.14 at
$E=6\times 10^{-11}$. This value is consistent with the magnitude of
the difference between the model and the numerical solution as shown in
Fig.~\ref{u_theo_prof1}b and Fig.~\ref{u_theo_prof2}. A similar result
has also been obtained with modes of higher order, and with modes of
the attractors listed in table~\ref{am}.

\begin{figure} \centering
   \begin{subfigure}{.3\textwidth} \vfill \centering
      \includegraphics[width=\linewidth,clip=true]{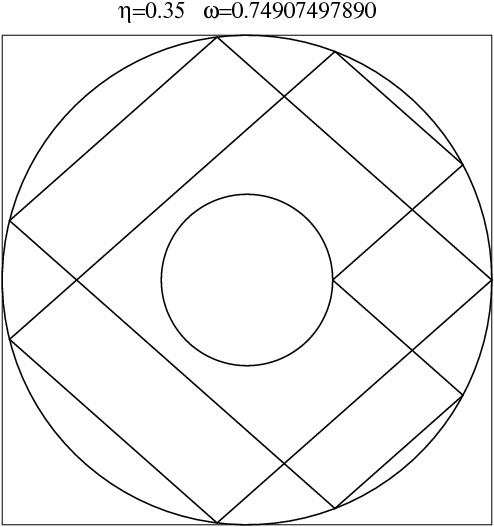}
      \caption[]{}
   \end{subfigure} \hfill
   \begin{subfigure}{.3\textwidth} \vfill \centering
\includegraphics[width=\linewidth,clip=true]{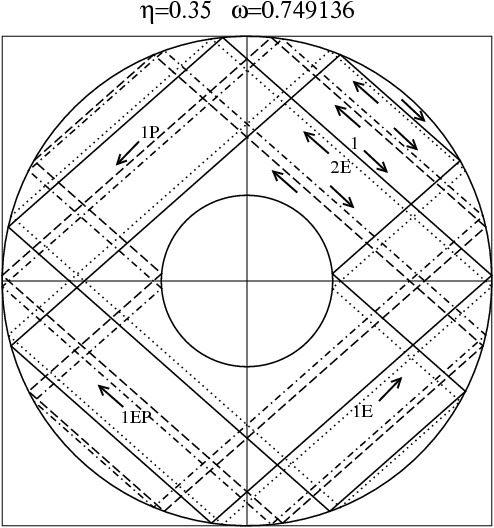}
      \caption[]{}
   \end{subfigure}
\begin{subfigure}{.3\textwidth} \centering
      \includegraphics[width=\linewidth,clip=true]{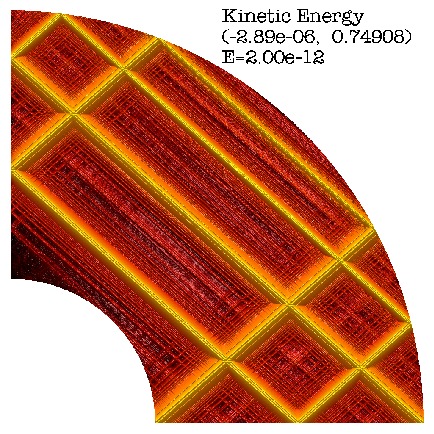}
\caption[]{}
   \end{subfigure}
\caption[]{(a) Shape of the asymptotic attractor at $\omega_0=0.7491$
and (b) its symmetrized view. Dashed, dotted, and dash-dotted lines
correspond to the symmetric
of the continuous lines with respectively rotation axis, equator and
origin. Arrows indicate the direction choosed along the attractor. 
(c) The corresponding numerical solution showing a meridional cut of the
kinetic energy density 
 (here $\eta=0.35$ and $E=2\times10^{-12}$).}
\label{attractor749}
\end{figure}

In figure \ref{u_theo_prof1}b, we also note that the actual attractor and
the asymptotic attractor both stand outside the shear layer. This is
because they are at distance $\calo{E^{1/6}}$ from $y_{\rm max}$, whereas
the shear layer width scales like $\calo{E^{1/4}}$.  As a consequence,
two shear layers adjacent to the asymptotic attractor do not ``see
each other" and seem to remain independent, unlike what happens in the
analogous two-dimensional problem analysed in \cite{RVG02}.

Hence, for a given eigenvalue the foregoing analysis gives a good
analytical approximation of the eigenfunction. Our procedure
however does not provide the quantization rule of eigenvalues observed numerically.

\subsection{Modes with $4n$-reflections on the rotation axis}

As we mentioned in the previous section, a special case occurs when the number $K$ of reflections on axis is 0 or
a multiple of 4.  In that situation, after a full loop along the attractor
the factor $(-i)^K$ due to the reflections on axis amounts to
unity.  Therefore the reflections on axis have overall no effect. Since
the governing reduced equation \eq{eq:reduced} is the same as the one
we obtained in the 2D toroidal configuration of \cite{RVG02}, the same
analysis should be valid here as well. We thus expect to find eigenvalues given
by the formula obtained in that paper:

\beq
\tautore_m=\pm (\omegatore_m-\omega_0)=-\big(m+\frac{1}{2}\big)\sqrt{\frac{\alpha_0 c E}{d}}, \quad m=0,1,2,...
\label{eq:disprel modes toriques}
\eeq
with eigenfunctions 

\beq
p_m=U(-m-1/2,z)=e^{-z^2/2}H_{m}(z), \quad
z=e^{-i\pi/8}(2\alpha_0 cd)^{1/4}\frac{E^{-1/4}y}{B}
\eeqn{parabolic_cylinder}
where $U$ is the parabolic cylinder function and
$H_m=(-1)^me^{z^2}d^m e^{-z^2}/dz^m$ are the Hermite polynomials.

We indeed found such modes. An example is the set of modes associated
with the attractor at $\omega_0=0.74907$ for $\eta=0.35$, shown in figure
\ref{attractor749}. We report in table \ref{749} the eigenvalues obtained
numerically together with the theoretical values given by \eq{eq:disprel
modes toriques}. There is a very good agreement between the difference of
consecutive eigenvalues and the spacing $|\tautore_{2m+2}-\tautore_{2m}|$
given by \eq{eq:disprel modes toriques}. However, the ratio
$\tau_1/\tautore_1$, between the observed and theoretical damping
rate of the fundamental mode is different from 1 and remains close
to the value  $\sim 1.5$ independently of the Ekman number and of the
attractor (see Tab.~\ref{749} for the 0.749 attractor). The reason for
this discrepancy is very likely due to the fact that we did not consider
the corrections induced by Ekman number to the reflection condition
on axis. Indeed, for the $4n$-attractors that do not cross the polar
axis (i.e. with $n=0$), the predicted eigenvalues and eigenfunctions
are perfectly verified (see figure \ref{fig:diamond} and table
\ref{table:diamond}).

\begin{figure} \centering
   \begin{subfigure}{.48\textwidth} \vfill \centering
      \includegraphics[width=\linewidth,clip=true]{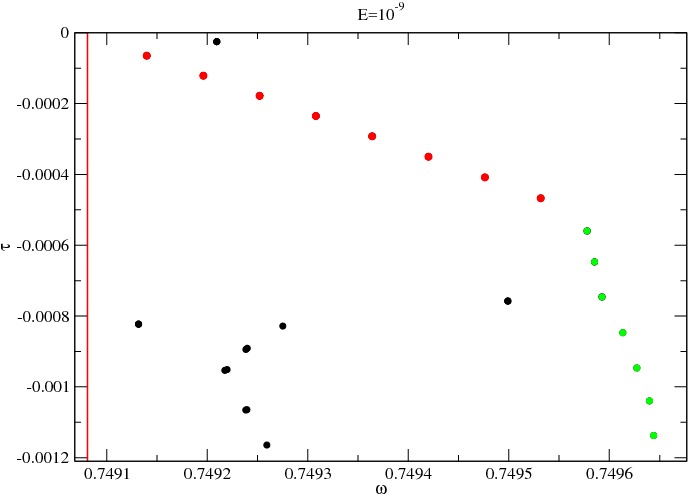}
      \caption[]{}
   \end{subfigure} \hfill
   \begin{subfigure}{.48\textwidth} \vfill \centering
      \includegraphics[width=\linewidth,clip=true]{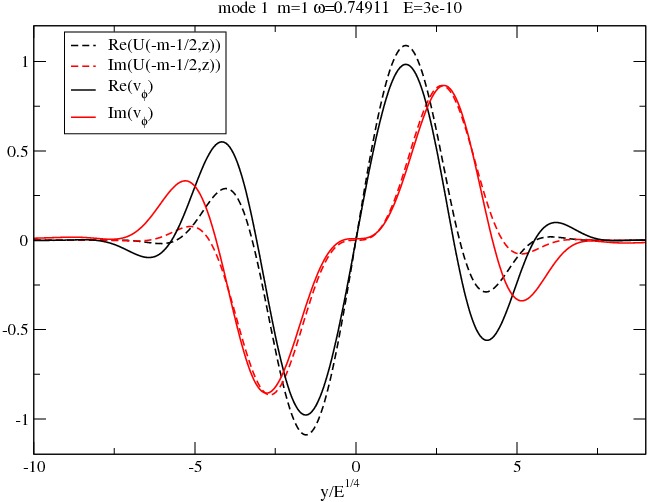}
      \caption[]{}
   \end{subfigure} \hfill
\caption[]{
(a) Spectrum of attractor $0.749$ for $E=10^{-10}$. The red
dots are the eigenvalues following the parabolic cylinder solution
\eq{parabolic_cylinder}.  The green eigenvalues follow the multi-scale
behaviour \eq{eq:u_theorique_revisited}.
(b) Profile of the eigenfunction corresponding to the least-damped of the
red eigenvalues of figure \ref{fig:0.749}a at $E=3\times 10^{-10}$.
}
\label{fig:0.749}
\end{figure}

\begin{table}
\tiny
\begin{center}
\begin{tabular}{c|c|c|c|c|c|c|c|c}
$n$ & $|\tau_n|$ & $|\omega_n-\omega_0|$  & $|\tau_{n+1}-\tau_n|$ & $|\omega_{n+1}-\omega_n|$ & $m$ & $|\tautore_m|$ & $|\tautore_{m+2}-\tautore_m|$ \\ \hline
\multicolumn{9}{c}{$E=3\times 10^{-10}$}\\
1 & $3.541\  10^{-5}$ & $3.544\  10^{-5}$ &                   &                   & 1 & $2.31\  10^{-5}$ &                 \\
2 & $6.610\  10^{-5}$ & $6.616\  10^{-5}$ & $3.069\  10^{-5}$ & $3.072\  10^{-5}$ & 3 & $5.38\  10^{-5}$ & $3.07\  10^{-5}$\\
3 & $9.700\  10^{-5}$ & $9.692\  10^{-5}$ & $3.090\  10^{-5}$ & $3.076\  10^{-5}$ & 5 & $8.45\  10^{-5}$ & $3.07\  10^{-5}$\\
4 & $1.280\  10^{-4}$ & $1.276\  10^{-4}$ & $3.104\  10^{-5}$ & $3.068\  10^{-5}$ & 7 & $1.15\  10^{-4}$ & $3.07\  10^{-5}$\\
5 & $1.592\  10^{-4}$ & $1.582\  10^{-4}$ & $3.112\  10^{-5}$ & $3.060\  10^{-5}$ & 9 & $1.46\  10^{-4}$ & $3.07\  10^{-5}$\\
\hline
\multicolumn{9}{c}{$E=2\times 10^{-12}$}\\
1 & 2.892e-06   & 2.892e-06   &             &             & 1 & 1.881e-06   & & \\
2 & 5.392e-06   & 5.392e-06   & 2.501e-06   & 2.5e-06     & 3 & 3.136e-06   & 2.509e-06   \\
3 & 7.894e-06   & 7.892e-06   & 2.502e-06   & 2.5e-06     & 5 & 4.39e-06    & 2.509e-06   \\
4 & 1.04e-05    & 1.039e-05   & 2.502e-06   & 2.5e-06     & 7 & 5.644e-06   & 2.509e-06   \\
\hline
\end{tabular}
\caption[]{First eigenvalues for the attractor $\omega_0=0.7491$,
$\eta=0.35$ at two Ekman numbers. The ratio
$\tau_1/\tautore_1\simeq 1.54$ seems to be independent of $E$.}
\label{749}
\end{center}
\end{table}

\begin{figure} \centering
   \begin{subfigure}{.26\textwidth} \vfill \centering
      \includegraphics[width=\linewidth,clip=true]{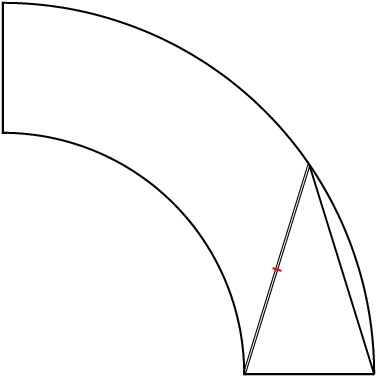}
      \caption[]{}
   \end{subfigure} \hfill
   \begin{subfigure}{.26\textwidth} \vfill \centering
\includegraphics[width=\linewidth,clip=true]{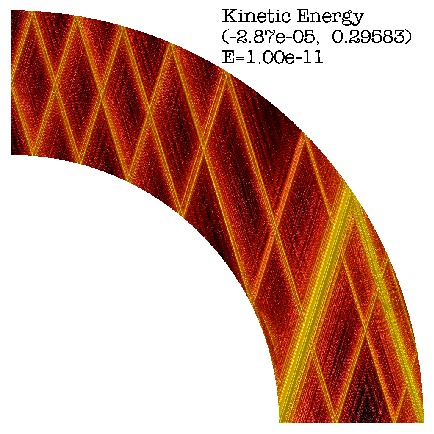}
      \caption[]{}
   \end{subfigure}
\hfill
\begin{subfigure}{.38\textwidth} \centering
      \includegraphics[width=\linewidth,clip=true]{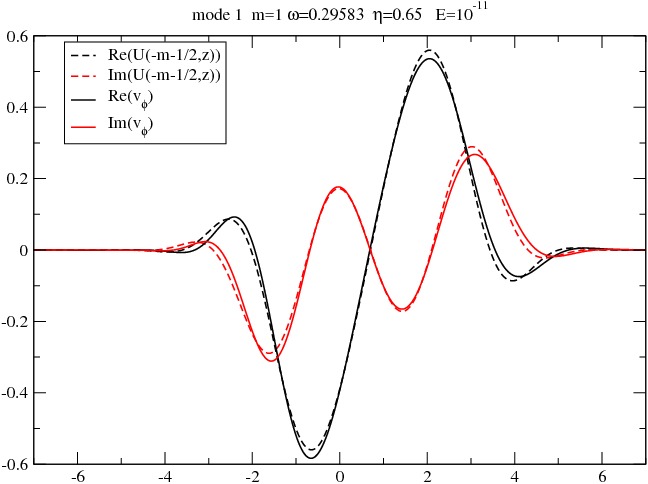}
\caption[]{}
   \end{subfigure}
\caption[]{
(a) Shape of the attractor at $\omega_0=0.2958$,
$\eta=0.65$ at Ekman number $E=10^{-11}$.
(b) The corresponding
numerical solution showing a meridional cut of the
kinetic energy density divided by the square root of the distance to the polar axis.
(c) Profile of the coresponding eigenfunction along the red line of panel (a). 
}
\label{fig:diamond}
\end{figure}

\begin{table}
\tiny
\begin{center}
\begin{tabular}{c|c|c|c|c|c|c|c|c}
$n$ & $|\tau_n|$ & $|\omega_n-\omega_0|$  & $|\tau_{n+1}-\tau_n|$ & $|\omega_{n+1}-\omega_n|$ & $m$ & $|\tautore_m|$ & $|\tautore_{m+2}-\tautore_m|$ \\ \hline
1 & 2.871e-05   & 2.866e-05   &             &             & 1 & 2.905e-05   & & \\ 
2 & 6.761e-05   & 6.739e-05   & 3.891e-05   & 3.874e-05   & 3 & 6.779e-05   & 3.874e-05   \\ 
3 & 1.066e-04   & 1.063e-04   & 3.899e-05   & 3.886e-05   & 5 & 1.065e-04   & 3.874e-05   \\ 
\hline
\end{tabular}
\caption[]{First eigenvalues for the attractor $\omega_0=0.2958$,
$\eta=0.65$ at Ekman number $E=10^{-11}$. Note that 
$\tau_1$ and $\tautore_1$ are almost equal.}
\label{table:diamond}
\end{center}
\end{table}

We also remark that even values of $m$ do not appear in the numerical
solution, but this can be explained by symmetry reasons as follows.
The numerical solutions are axisymmetric (see section 2), but they are
also symmetric with respect to the equator (see beginning of section
3). On the other hand the asymptotic attractor is symmetric with respect
to the equator (see figure \ref{attractor749}) but the actual attractor
(at $\omega\neq\omega_0$) does not have any symmetry. The analytical
solution \eq{parabolic_cylinder} for the shear layer along the attractor
must therefore be symmetrized in order to fulfill the symmetries imposed to the
numerical solution.  This is done by adding replicas of the solution
along the asymptotic attractor suitably symmetrised with respect to the
original one: starting from the attractor denoted with continuous lines
in figure \ref{attractor749}, we construct three additional attractors:
the first is obtained through the axial symmetry (the dashed lines),
the second through the equatorial symmetry (the dotted lines) and
the third through the combined axial and equatorial symmetry (the
dash-dotted lines).  The set of these four attractors makes the figure
symmetric with respect to both the rotation axis and the equator, 
as numerically required.
For clarity in each panel we label with 1, 1E, 1A and 1EA a given branch
of each attractor. The numerical solution is expected to be
the same along each of these branches.

The two neighbouring branches marked by 1 and 2E in figure
\ref{attractor749} form the shear layer whose profile is shown in figure
\ref{fig:0.749}a. The large amplitude negative values of $y$ belong
to the branch 2E while the positive values of $y$ correspond to the
branch 1. It is readily seen that $v_\phi(y)$ must be the opposite of
$v_\phi(-y)$ because the reflection on the outer sphere connecting branch
1E to branch 2E produces a change of sign on $v_\phi$.  The analytical
solutions \eq{parabolic_cylinder} with even values of $m$ however are even
functions of $y$: $v_\phi(-y)=v_\phi(y)$ and are thus forbidden. Those
with $m$ odd on the other hand are odd functions of $y$ and are thus
allowed, as we observe numerically.  The shape of the least-damped
mode ($n=1$) is indeed similar to the predicted one for $m=1$, as we
can see in figure \ref{fig:0.749}a.  Here we have considered the
0.749 attractor but the reasoning and the conclusions are valid for all
the asymptotic attractors that are equatorially symmetric and have $4n$
reflections on axis. We finally remark that if we had solved numerically
the equatorially antisymmetric problem, we would have obtained to good
precision the 2D eigenmodes corresponding to even $m$ values.

\begin{figure} \centering
   \begin{subfigure}{.3\textwidth} \vfill \centering
      \includegraphics[width=\linewidth,clip=true]{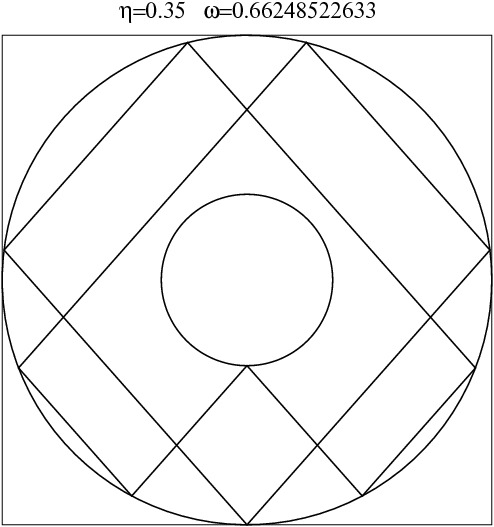}
      \caption[]{}
   \end{subfigure} \hfill
   \begin{subfigure}{.3\textwidth} \vfill \centering
\includegraphics[width=\linewidth,clip=true]{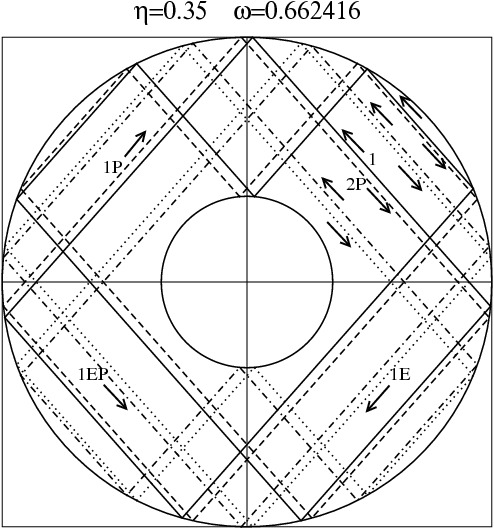}
      \caption[]{}
   \end{subfigure}
\begin{subfigure}{.3\textwidth} \centering
      \includegraphics[width=\linewidth,clip=true]{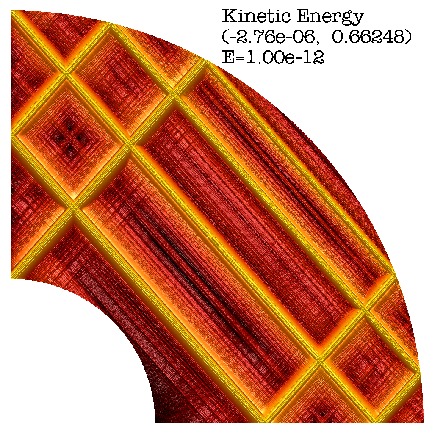}
\caption[]{}
   \end{subfigure}
\caption[]{(a) Shape of the asymptotic attractor at $\omega_0=0.6625$
and (b) its symmetrized view. Dashed, dotted, and dash-dotted lines
correspond to the symmetric
of the continuous lines with respectively rotation axis, equator and
origin. Arrows indicate the direction choosed along the attractor. 
(c) The corresponding numerical solution showing a meridional cut of the
kinetic energy density 
 (here $\eta=0.35$ and $E=\times10^{-12}$).}
\label{attractor662}
\end{figure}

\begin{table}
\begin{center}
\begin{tabular}{c|c|c|c|c|c|c|c|c|c}
$E$ & $n$ & $|\tau_n|$ & $|\omega_n-\omega_0|$  & $|\tau_{n+1}-\tau_n|$ & $|\omega_{n+1}-\omega_n|$ & $m$ & $|\tautore_m|$ & $|\tautore_{m+2}-\tautore_m|$ \\ \hline
$10^{-10}$& 1 & $1.287\  10^{-5}$ & $0.789\  10^{-5}$ &                   &                   & 0 & $0.47\  10^{-5}$ &                 \\
$10^{-10}$& 2 & $3.418\  10^{-5}$ & $2.928\  10^{-5}$ & $2.131\  10^{-5}$ & $2.139\  10^{-5}$ & 2 & $2.35\  10^{-5}$ & $1.88\  10^{-5}$\\
$10^{-10}$& 3 & $5.454\  10^{-5}$ & $4.935\  10^{-5}$ & $2.036\  10^{-5}$ & $2.007\  10^{-5}$ & 4 & $4.23\  10^{-5}$ & $1.88\  10^{-5}$\\
$10^{-10}$& 4 & $7.453\  10^{-5}$ & $6.902\  10^{-5}$ & $1.999\  10^{-5}$ & $1.967\  10^{-5}$ & 6 & $6.10\  10^{-5}$ & $1.88\  10^{-5}$\\
\hline
$10^{-12}$& 2 & $2.759\  10^{-6}$ & $2.288\  10^{-6}$ &                   &                   & 2 & $2.35\  10^{-6}$ &                 \\
$10^{-12}$& 3 & $4.766\  10^{-6}$ & $4.332\  10^{-6}$ & $2.006\  10^{-6}$ & $2.043\  10^{-6}$ & 4 & $4.23\  10^{-6}$ & $1.88\  10^{-6}$\\
\hline
\end{tabular}
\caption[]{First eigenvalues for attractor $\omega_0=0.66249$, $\eta=0.35$. }
\label{662}
\end{center}
\end{table}

\begin{figure} \centering
   \begin{subfigure}{.48\textwidth} \vfill \centering
      \includegraphics[width=\linewidth,clip=true]{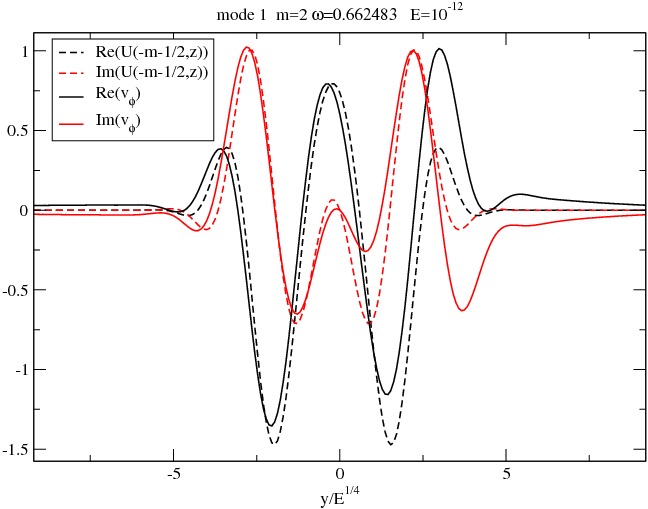}
      \caption[]{}
   \end{subfigure}
   \begin{subfigure}{.48\textwidth} \vfill \centering
      \includegraphics[width=\linewidth,clip=true]{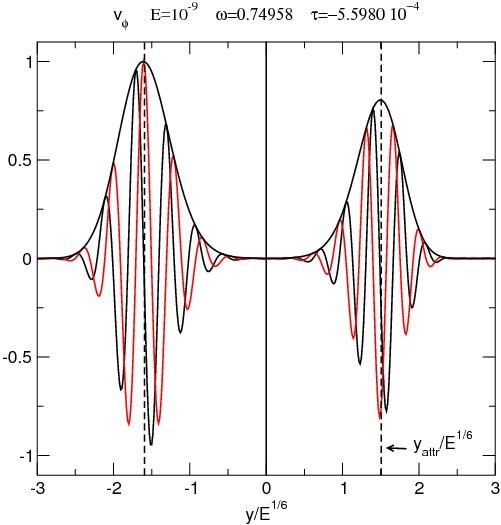}
      \caption[]{}
   \end{subfigure}
\caption[]{
(a) Profile of the eigenfunction corresponding to the least-damped of
the eigenvalues of the attractor $0.66249$ at $E=10^{-12}$.
(b) Profile of the eigenfunction corresponding to the least-damped of
the green eigenvalues of spectrum in figure~\ref{fig:0.749}.
}
\label{fig:0.662}
\end{figure}

Let us now turn to the case where the asymptotic attractor has $4n$
reflections on axis but no equatorial symmetry, like the attractor
plotted in figure \ref{attractor662}. One such attractor is obtained
by rotating clockwise the attractor of figure~\ref{attractor749}a
by $\pi/2$; its frequency is $\omega_0=0.66249=\sqrt{1-0.74907^2}$.
We observed that the eigenvalue associated with this attractor are
loosely related to those given by \eq{eq:disprel modes toriques}.
The matching between analytics and numerics is much worse than for the
previous attractor.  The difference also shows up in the eigenfunction:
figure \ref{fig:0.662}a shows the numerical velocity profile together
with the analytic prediction \eq{parabolic_cylinder} for mode $n=2$
at $E=10^{-12}$. Despite a very low value of the Ekman number, the two
functions still show noticeable differences.This mismatch is due to
symmetry requirements of the numerical solution that cannot be satisfied
by the analytical solution \eq{parabolic_cylinder}.  To show this, we
first symmetrize the attractor in order to respect the symmetries imposed
to the numerical solution (solutions must be axisymmetric and equatorially
symmetric). The layout of the attractors after symmetrization is shown
in figure \ref{attractor662}b.  In this figure the two neighbouring
branches marked ``1" and ``2P" form the shear layer whose profile is shown
in figure \ref{fig:0.662}a.  Branch 2P is the continuation of branch 1P,
and branch 1P is the mirror symmetric of branch 1.  Branch 2P is reached
from branch 1P after reflection on outer boundary and rotation axis.
Reflection in outer boundary produces a change of sign, and crossing
of rotation axis produces a $-i$-factor. We get thus the condition
$v_\phi(-y)=iv_\phi(y)$. This relation is however not satisfied by any of
the functions \eq{parabolic_cylinder}.  So there cannot be solutions of
the type \eq{parabolic_cylinder} as $E\to 0$. This impossibility likely
explains why the $n=1$-mode of the $\omega_0=0.66249$-attractor has
a frequency that goes out of the range of existence of the attractor
when $E\leq 2\times 10^{-12}$.  The same considerations hold for
all the asymptotic attractors that are symmetric with respect to the
polar axis with reflections on axis that are multiple of 4: solutions
\eq{parabolic_cylinder} are not expected to exist asymptotically for
these modes.

We finally note that the shear layer analysis 
of section 4 
is still valid and so we expect to obtain modes described by
formula \eq{eq:u_theorique_revisited}. Indeed, eigenvalues marked
in green in figure \ref{fig:0.749} correspond to such a case. We show
in figure \ref{fig:0.662}(b) the profile of the least damped of these
modes and remark that the position of $\ymax$ and $\yattr$ coincide, which
is consistent with \eq{ymax/yattr} with $K=0$.

\section{Conclusions}

In this work we continued our investigations of the properties of inertial
modes in a spherical shell started in \cite{RV97}, \cite{RGV01} and
\cite{RVG02}. The possibility of using more computing power or enhanced
precision, allowed us to establish a simple mathematical law \eq{egv_exp}
for the eigenvalues of the modes that are associated with
some attractors made of a periodic orbit of characteristics. For these
modes, we identified three scales that determine the structure of the
shear layers constituting the eigenmodes. These scales are controlled by
fractional powers of the Ekman number, namely $E^{1/6}$, $E^{1/4}$ and
$E^{1/3}$. They singularize the small parameter $E^{1/12}$. This very
low power of the Ekman number shows that the true asymptotic regime,
such that $E^{1/12}\ll1$, is not reachable by numerical solutions. It
may not even be relevant to the extremely low Ekman numbers met in
astrophysics that can hardly go below $10^{-18}$. However,  it remains
interesting to understand the structure of the solutions when the Ekman
number is very small, yet finite.

The present limits of numerical solutions are no longer the available
memory, which controls the reachable spatial resolution, but the
round-off errors boosted by the ill-conditioned operator.
This ill-conditioning is related to the singular nature of
the inviscid limit of eigenfunctions. It may be circumvented by
using enhanced
precision. 
We therefore
put our effort on converting our code to use extended precision
(quadruple precision). This choice rapidly reached however the limits
of present technology, since computers are all built with double
precision arithmetics. Extended arithmetics is therefore obtained
through software programming and is thus very slow. It limited our
calculations to Ekman numbers above $10^{-9}$. Nevertheless, high order
modes, which are very sensitive to round-off errors
could be properly computed \cite[see also][]{VRBF07}.

The foregoing numerical results obtained on the ``attractor modes"
guided our analysis of their structure and thanks to the reduced problem
\eq{eq:reduced}
simplifying the original equation we could determine an analytic
formula for the shape of the shear layers. It turns out that an
attractor mode is a wave trapped around a characteristic attractor whose
typical wavelength is \od{E^{1/3}} but whose envelope has a width
\od{E^{1/4}}. This wave packet remains at a distance \od{E^{1/6}} from the
asymptotic attractor that has a vanishing Lyapunov exponent.  Our analysis
does not provide a selection rule for the eigenvalues.  It is most likely
that the simplifications we made to retrieve the structure of the shear
layer are too strong to allow for the determination of the quantization
rule of the modes. Our analysis indeed was restricted to the $E^{1/3}$
and $E^{1/4}$ scales and did not include the $E^{1/6}$ one: the condition
leading to the quantization of the solution might be more deeply nested
in the multiscale dependence of the solutions.  The special case where
attractors have $4n$ reflections on the rotation axis has interestingly
extended the applicability of the 2D-model solved by \cite{RVG02}.
However, the predicting power of the 2D-model is limited to the
frequency spacing of some modes verifying some given symmetries. Here
too, some piece seems to be missing for the model to make accurate
predictions of eigenvalues and eigenmodes.

Beside attractor modes, we also got evidence of the existence of
critical latitude modes. These modes are made of detached shear layers
emitted by the critical latitude singularity on the inner boundary. They
connect the northern and southern critical latitude singularities. Since
the path of characteristics from one singularity to that of the other
hemisphere is not unique, this set of modes is determined by the set of
paths and the transverse wave number of the shear layer. The inspection
of their damping rates, in the range $10^{-9}\infapp E \infapp 10^{-7}$,
shows a dependence with the Ekman number close to $E^{0.8}$, meaning a
weak dependence of the width of the layers with this number. However,
this behaviour does not seem to be asymptotic, since it disappears when
$E\infapp 10^{-9}$ for the least-damped mode. More work is needed to
fully understand the behaviour of these modes at lower viscosities.

The last category of modes that we identified are a series of
modes whose frequency is close to $\sin(\pi/4)$. We recall that when
$\omega=\sin(\pi/4)$ characteristics follow strictly periodic orbits and
thus no small scale is forced by the mapping \cite[][]{RGV01}. The modes
of this kind seem to be essentially inertial waves trapped between the
two shells.  We qualified them as quasi-regular modes because in some
range of Ekman numbers they behave as regular modes: their eigenfunction
is almost independent of $E$. However, this does not mean that they
exist in the inviscid limit. We find that the modes follow quite simple
quantization rules, which show that when $\omega\tv\sin(\pi/4)$, the
typical wavenumber of the mode tends to infinity and so does the damping
rate. This result explains the no-response flow of a fluid in a rotating
shell when it is forced periodically at $\omega=\sin(\pi/4)$, as has been
observed by \cite{RV10}. We expect that such a phenomenon occurs for
all frequencies leading to neutral periodic orbits of characteristics.
With our set-up (aspect ratio $\eta=0.35$), this should also occur
when $\omega=\sin(\pi/6)$ and $\omega=\sin(\pi/3)$, but inspection of
the modes around these frequencies does not show a neat quasi-regular
behaviour. Perturbations from the critical latitude singularity appear
to be important.  For the aspect ratio $\eta=0.35$ it seems that
only $\sin(\pi/4)$ can produce quasi-regular modes, but it turns out
that $\omega=\sin(\pi/8)$, for a smaller core ($\eta=0.20$), actually
owns also quasi-regular modes. Hence, beside the neutral character of
periodic orbits, some other virtue (to be uncovered) is needed to allow
quasi-regular modes.  Finally, the regularity of the modes, which we
characterize by the proportionality of the damping rate to the Ekman
number, is lost when the Ekman number is below some value specific to
the mode. Thin shear layers appear and introduce a stronger dissipation.

The foregoing solutions, although derived in a highly idealized set-up
show the extreme richness and complexity of the dynamics of rotating
fluids. The oscillation spectrum of an incompressible slightly viscous
fluid inside a rotating spherical shell appear much more complex than
our first studies \cite[][]{RV97} let us think. We now clearly see that
the eigenvalues cannot be represented by a single formula. Because of
the very small powers of the Ekman number ($E^{1/12}$) that seem to
control the eigenmodes around attractors, even the astrophysical regime
is not in the asymptotic state of vanishingly small quantities.We face
here the same difficulty as \cite{SLD13} when they studied the
libration-induced flows in a spherical shell.

Hence, the asymptotic spectrum at vanishing (but non-zero) Ekman number
is most probably a composition of different sets of eigenvalues, which
follow their own asymptotic laws. More work is still needed to exhibit
the analytical solutions that describe this asymptotic limit like the one
obtained by \cite{RVG02} on the 2D problem.

Back to astrophysics and geophysics, which motivate these
investigations (since the work of \citealt{poinc1885}),
the various sets of modes and their different asymptotic behaviour will
impact the response of the fluid to a global forcing like a tidal one.
Stars and planets are fairly more complicated systems than our simple
spherical shell, but this system has pointed out mechanisms that may
persist when stratification or differential rotation (or
both) are taken into account \cite[e.g.][]{DRV99,MBRB16}.

\begin{acknowledgements}
This work was performed using HPC resources from CALMIP (Grant
2016-07). We are especially grateful to P. Barbaresco and N. Renon for
their assistance in running the very RAM-demanding jobs needed to explore
the very low Ekman number space. MR acknowledges the support of ISSI
(programme on the "Seismology of Fast Rotating Stars", PI J. Ballot)
for allowing fruitful discussions on the results of the present work.
\end{acknowledgements}

\bibliography{../../../biblio/bibnew}
\appendix

\section{Derivation of the reduced equation}

We start from the viscous Poincaré equation \eq{poincviscb} which we
rewrite using the $(x,y)$ coordinates in the meridian plane instead of
$(r,z)$. We note that 

\[ \Delta' = \ddx{}+\ddy{}+\frac{1}{4s^2}\]
so that this operator may be reduced to $\partial^2/\partial y^2$ since
we are considering the solution associated with thin shear layers. Thus
doing we reduce \eq{poincviscb} to

\beq
\lp\lambda-E\ddy{}\rp^{\!\!2}\ddy{p}+\omega^2\ddy{p}+2\alpha\omega\dxdy{p} = 0
\eeq
Now focusing on solutions with smallest $y$-scales $E^{1/3}$, the term
with largest derivative $-E\partial^6/\partial y^6$ is negligible. Integrating
over $y$ and noting that $\lambda = \tau+ i\omega$ we get:

\beq
\alpha \dx{p} = \frac{\lambda E }{\omega} \frac{\partial^3 p}{\partial y^3} + i \tau \dy{p}
\eeqn{A2}

We can safely replace $\alpha$ with the asymptotic value $\alpha_0$
since their difference is $\od{E^{1/3}}$ thus negligible.  In this
equation the smallest scale $E^{1/3}$ makes all the terms of the same
order ($\tau$ is of order $E^{1/3}$). However, the equation remains valid
if, as seen numerically, larger scales and/or contributions to $\tau$
of order $E^{1/2}$ are retained.

%
%

\section{Reflection on the rotation axis}

We show that, for the inviscid case, the reflection on 
axis produces a $-i$ factor in the
solution. Consider the inviscid axisymmetric solution propagating
in the direction of increasing $z$, namely

\[ p(s, z, t) = p(s)e^{i(kz-\omega t)}\]
Poincar\'e equation yields

\[
\ddpptwo{p}{s}+\frac{1}{s}\ddpp{p}{s}+\frac{\alpha^2 k^2}{\omega^2}p=0
\]
which is solved by

\[ p(s) = AJ_0(k_ss)\]
where $k_s = \alpha k/\omega$ and $J_0$ is the zeroth order Bessel
function. Asymptotically, when $k_ss\gg1$, namely far from the rotation
axis

\[J_0(k_ss)\simeq\sqrt{\frac{2}{\pi k_s s}} \cos\left(k_ss-\frac{\pi}{4}\right)
=\sqrt{\frac{2}{\pi k_s s}} \left(e^{ik_ss-i\pi/4}+e^{-ik_ss+i\pi/4}\right)\]
which shows that the outward wave 

\[ e^{ik_ss-i\omega t -i\pi/4}\]
is shifted by $-i\pi/2$ compared to the inward wave

\[ e^{-ik_ss-i\omega t +i\pi/4}\]
if we consider the propagation in a $z=\cst$ plane.  We thus conclude
that the bounce of the wave on the axis, ingoing and then outgoing,
imprints a factor $-i$ to the solution. This behaviour also emerges in
the analysis of shear layers produced by a librating disc, which reflect
on the axis \cite[e.g.][]{ledizes15}.

\end{document}